\documentclass[12pt,preprint]{aastex}

\newcommand{\cldy}{{\it Cloudy}}

\newcommand{\gtsimeq}{\raisebox{-0.6ex}{$\,\stackrel
        {\raisebox{-.2ex}{$\textstyle >$}}{\sim}\,$}}

\received{}
\revised{}
\accepted{}

\shorttitle{V1186 Sco}
\shortauthors{Schwarz et al.}

\begin{document}

\title{The Early Spectrophotometric Evolution of V1186 Scorpii (Nova Scorpii 2004 \#1)}

\author{
G.~J. Schwarz\altaffilmark{1},
C.~E. Woodward\altaffilmark{2},
M.~F. Bode\altaffilmark{3}, 
A. Evans\altaffilmark{4},
S.~P. Eyres\altaffilmark{5}, 
T.~R. Geballe\altaffilmark{6}
R.~D. Gehrz\altaffilmark{2}, 
M.~A. Greenhouse\altaffilmark{7},
L.~A. Helton\altaffilmark{2},
W. Liller\altaffilmark{8},
J.~E. Lyke\altaffilmark{9},
D.~K. Lynch\altaffilmark{10,18},
T.~J. O'Brien\altaffilmark{11},
R.~J. Rudy\altaffilmark{10,18},
R.~W. Russell\altaffilmark{10,18},
S.~N. Shore\altaffilmark{12}, 
S.~G. Starrfield\altaffilmark{13},
T. Temim\altaffilmark{2},
J.~W. Truran\altaffilmark{14},
C.~C. Venturini\altaffilmark{10}, 
R.~M. Wagner\altaffilmark{15}
R.~E. Williams\altaffilmark{16},
R. Zamanov\altaffilmark{3,17}
}


\altaffiltext{1}{West Chester University, Department of Geology and Astronomy,
750 S. Church St., West Chester, PA 19383,\ {\it gschwarz@pha.jhu.edu}} 

\altaffiltext{2}{Department of Astronomy, School of Physics and Astronomy, 
116 Church Street, S.~E., University of Minnesota, Minneapolis, MN 55455} 

\altaffiltext{3}{Astrophysics Research Institute, Liverpool 
John Moores University, Twelve Quays House, Birkenhead CH41 1LD, UK}

\altaffiltext{4}{Astrophysics Group, Keele University, Keele, 
Staffordshire ST5, 5BG, UK} 

\altaffiltext{5}{Department of Physics, Astronomy and Mathematics,
University of Central Lancashire, Preston PR1 2HE, UK}

\altaffiltext{6}{Gemini Observatory, 670 North A'ohoku Place, 
Hilo, HI 96720}

\altaffiltext{7}{NASA Goddard Space Flight Center, Code 665,
Greenbelt, MD 20771}


\altaffiltext{8}{Institute for Nova Studies, Casilla 5022, Vi\~{n}a del Mar, 
Chile}

\altaffiltext{9}{W.~M. Keck Observatory, 65-1120 Mamalahoa Hwy., Kamuela,
HI 96743}

\altaffiltext{10}{The Aerospace Corporation, Mail Stop 2-266, P.O. Box
92957, Los Angeles, CA 90009-2957}

\altaffiltext{11}{Jodrell Bank Observatory, University of Manchester,
Macclesfield Cheshire SK11 9DL, UK}


\altaffiltext{12}{Dipartimento di Fisica ``Enrico Fermi,'' Universita' di 
Pisa, largo Pontecorvo 3, Pisa 56127 Italy; INFN - Sezione di Pisa} 

\altaffiltext{13}{School of Earth and Space Exploration, Arizona State 
University, P.O. Box 871404, Tempe, AZ 85287} 

\altaffiltext{14}{Department of Astronomy and Astrophysics, University of 
Chicago, 5640 S. Ellis Avenue, Chicago, IL 60637 and Argonne National 
Laboratory, 9700 South Cass Road, Argonne, IL 60439} 

\altaffiltext{15}{Large Binocular Telescope Observatory, University
of Arizona, 933 North Cherry Avenue, Tucson, AZ 85721}

\altaffiltext{16}{Space Telescope Science Institute, 3700 San Martin 
Drive, Baltimore, MD 21218}

\altaffiltext{17}{Bulgarian National Astronomy Institute}

\altaffiltext{18}{Visiting Astronomer at the Infrared Telescope Facility,
which is operated by the University of Hawaii under Cooperative
Agreement NCC 5-538 with NASA Office of Space Science, Planetary
Astronomy Program}



\begin{abstract}

We report optical photometry and optical through mid-infrared spectroscopy
of the classical nova V1186 Sco.  This slowly developing nova had an 
complex light curve with multiple secondary peaks similar to those seen
in PW Vul.  The time to decline 2 magnitudes, t$_2$, was 20 days but the 
erratic nature of the light curve makes determination of intrinsic 
properties based on the decline time (e.g., luminosity) problematic, 
and the often cited MMRD relationship of \citet{DVL95} fails to yield a 
plausible distance. Spectra covering 0.35 to 35~$\mu$m were obtained in 
two separate epochs during the first year of outburst. The first 
set of spectra, taken about 2 months after visible maximum, are typical 
of a CO-type nova with narrow 
line emission from \ion{H}{1}, \ion{Fe}{2}, \ion{O}{1} and \ion{He}{1}. 
Later data, obtained between 260 and 380 days after maximum, reveal an
emerging nebular spectrum. \textit{Spitzer} spectra show  
weakening hydrogen recombination emission with the emergence of 
[\ion{Ne}{2}] (12.81~$\mu$m) as the strongest line. Strong emission 
from [\ion{Ne}{3}] (15.56~$\mu$m) is also detected. Photoionization 
models with low effective temperature sources and only marginal neon 
enhancement (Ne $\sim$ 1.3~Ne$_{\odot}$) are consistent with these 
IR fine-structure neon lines indicating that V1186 Sco did not occur 
on a ONeMg white dwarf.  In contrast, the slow and erratic light curve 
evolution, spectral development, and photoionization analysis of the ejecta 
imply the outburst occurred on a low mass CO white dwarf. We note that
this is the first time strong [\ion{Ne}{2}] lines have been detected
so early in the outburst of a CO nova and suggests that the presence 
of mid-infrared neon lines is not directly indicative of a 
ONeMg nova event.

\end{abstract}


\section{INTRODUCTION}
\label{sec:intro}

Classical nova outbursts occur in binaries consisting of a white dwarf 
(WD) and a low mass star that has filled its Roche lobe.  Matter is 
accreted onto the WD with the deeper layers becoming degenerate.  The 
temperature increases, first through compression and then rapidly when 
thermonuclear reactions begin, eventually leading to a thermonuclear 
runaway.  Energy deposited in the accreted mass is sufficient to eject a 
fraction of the material.  The amount of mass accreted and 
subsequently ejected, along with the energetics of the outburst are 
related to the mass and composition of the underlying WD.  Outbursts on 
Carbon-Oxygen (CO) WDs tend to eject more mass under less energetic 
conditions than novae on Oxygen-Neon-Magnesium (ONeMg) WDs
\citep{Schwarz97,Schwarz01,Schwarz07}. 

\object{V1186 Sco} was discovered on 03.1 July 2004~UT by \citet{IAUC8369} 
at V = 11.98.  The earliest optical spectrum of V1186 Sco was obtained on 
06.51 July 2004~UT, or about 3 days before visual maximum \citep{Fujii04}. 
This spectrum displayed strong Balmer emission lines with FWHM 
of order 500~km~s$^{-1}$ and possible P-Cygni profiles, bearing a strong 
resemblance to the pre-maximum spectra of the slow nova \object{V723 Cas} 
\citep{Evans03,Munari96}.  \textit{Spitzer Space Telescope} Target of 
Opportunity (ToO) observations were immediately activated and V1186 Sco was 
observed during two observational epochs in the first year of outburst.  
The \textit{Spitzer} data were complemented with contemporaneous optical and 
near-IR (NIR) spectra obtained as part of a coordinated multi-wavelength 
nova observing initiative.  These data provided wavelength coverage from 0.35
to 35 $\mu$m with a variety of spectral resolutions. In addition, the early 
photometric light curve is well sampled showing a 
leisurely rise and subsequent decline (see \S2.1).  The light curve behavior, 
the low expansion velocities, and the similarity of the early spectra of 
V1186 Sco \citep[][and \S2.2]{Fujii04} to those of V723 Cas \citep{Evans03} 
and \object{PW Vul} \citep{Gehrzet88,Evans90,Will96} implies that 
the outburst occurred on a low mass CO-type WD.  The behavior of V1186 Sco 
contrasts markedly with that of the other nova discovered in the constellation
Scorpius in 2004, the fast ONeMg nova \object{V1187 Sco} \citep{lynch2006}.
In particular, the development of infrared coronal lines detected with 
\textit{Spitzer} was significantly more rapid in V1187 Sco (cf., \S4).  

We present the photometric and spectroscopic data in \S2. 
Section 3 discusses the estimate of the reddening and distance to V1186 
Sco. Section 4 describes the line evolution, particularly the significant
detection of [\ion{Ne}{2}] and [\ion{Ne}{3}] in the \textit{Spitzer}
spectra. A photoionization model of the later multiwavelength spectra
is also presented in \S4.  Section 5 presents a discussion of our primary
results followed by our conclusions in \S6.

\section{OBSERVATIONS}
\label{sec:obs}

\subsection{Photometry}

\begin{figure}
\plottwo{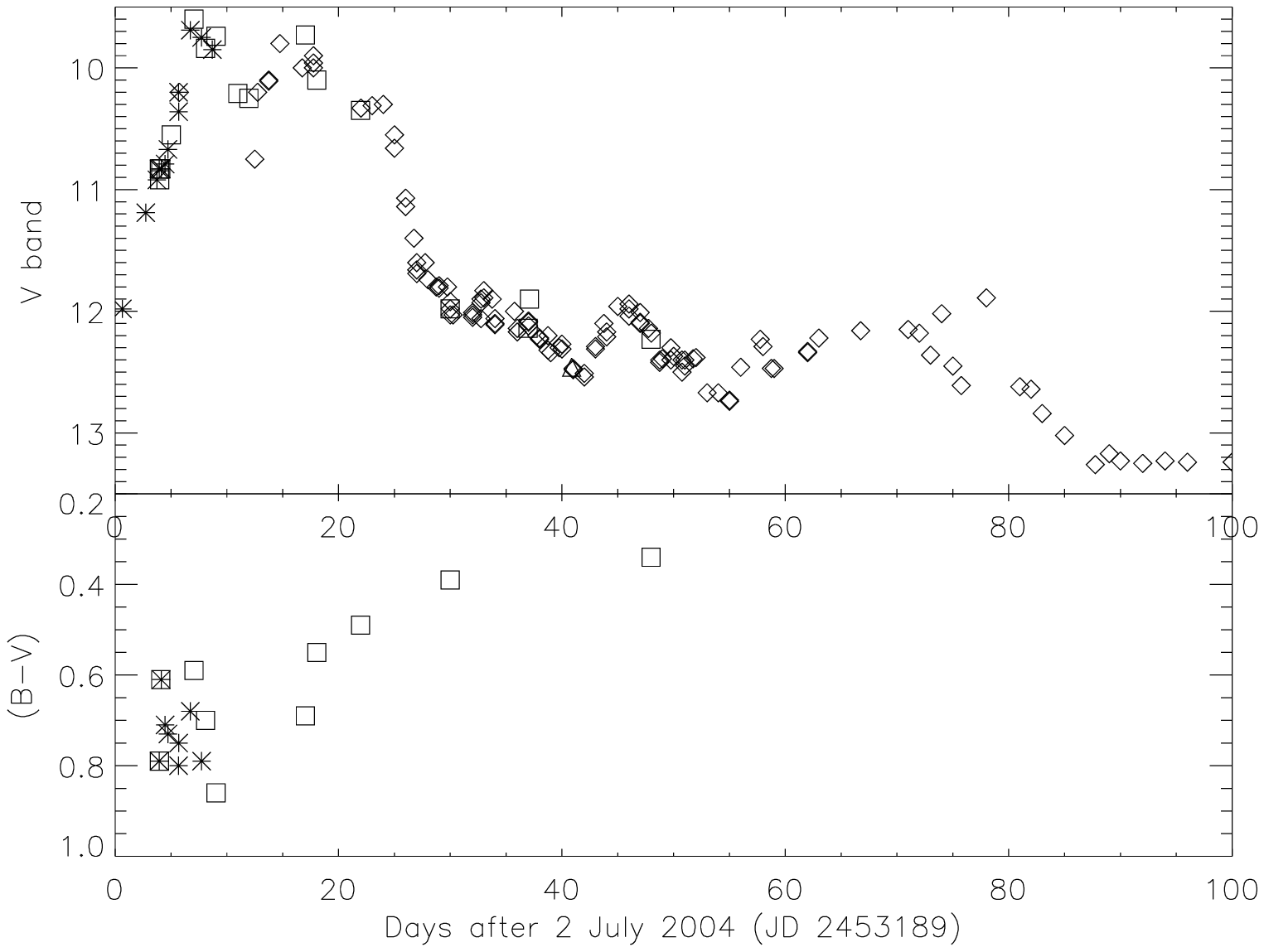}{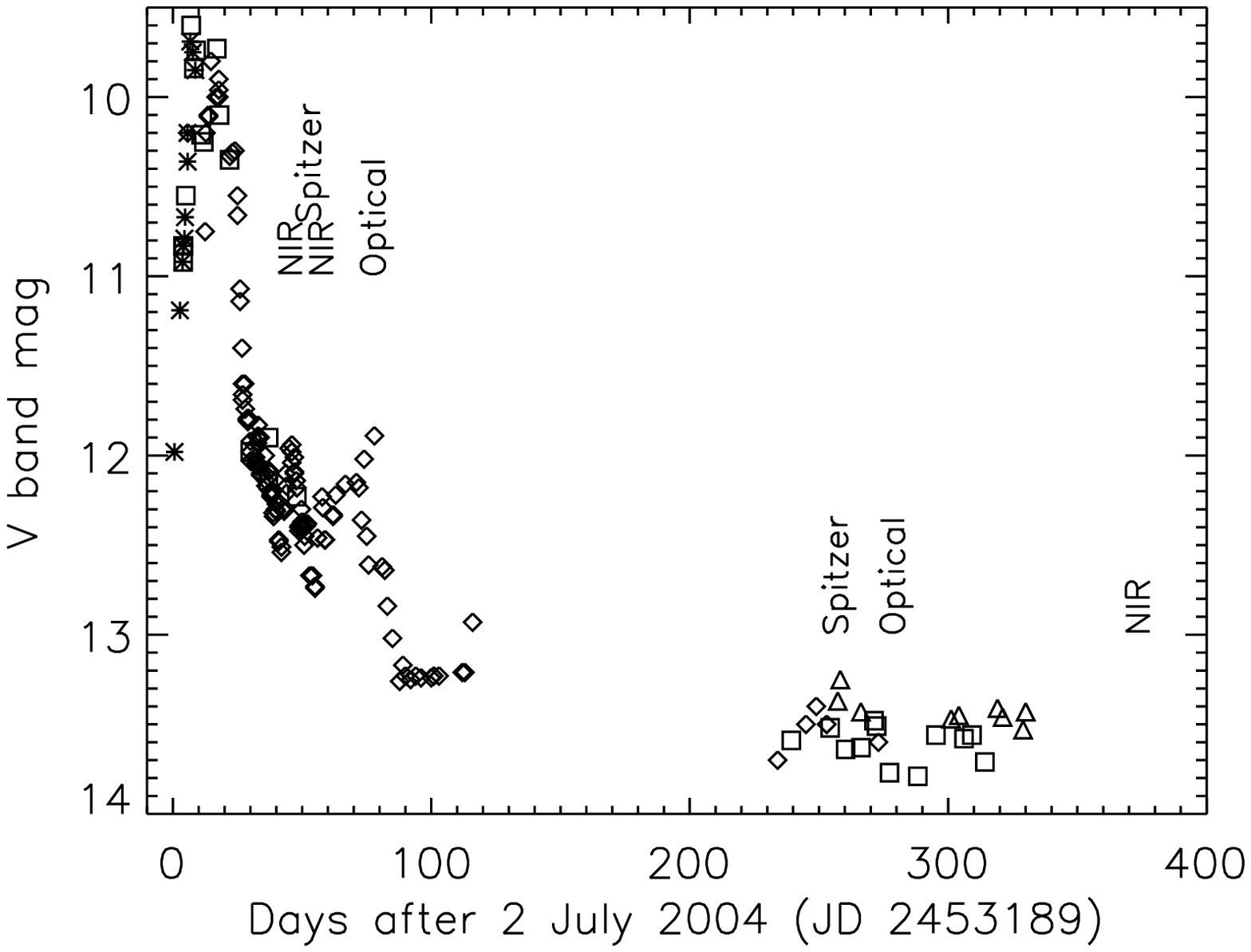}
\caption{The light curve of V1186 Sco.  The V band photometry was taken 
from VSNET ({\it squares}), IAU Circulars ({\it asterisks}), 
AAVSO ({\it diamonds}), and our own Liverpool 2-m data ({\it triangles}). 
The {\it left} plot shows the early evolution and the {\it right} 
figure shows the light curve over the first year. The lower portion of 
the {\it left} figure shows the ($B-V$) 
color evolution from the available data.
\label{aavso-bb}}
\end{figure}

The early light curve of V1186 Sco is shown in Fig.~\ref{aavso-bb}a, 
and the first year light curve is presented in Fig.~\ref{aavso-bb}b. 
The V band data are from the
IAU circulars, VSNET, the AAVSO \citep{aavso} and the 
Liverpool Robotic Telescope \citep{steele04}\footnote{The 2-m fully robotic 
Liverpool Telescope is operated on the island of La Palma by Liverpool John 
Moores University in the Spanish Observatorio del Roque de los Muchachos of 
the Instituto de Astrofisica de Canarias.}. Photometric data were obtained 
using the $2048 \times 2048$ CCD camera 
``RATCam''\footnote{\url{http://telescope.livjm.ac.uk}}. Photometric 
data were reduced using the GAIA image analysis software package and are 
provided in Table~\ref{tab:lrt-phot}. 

The V1186 Sco light curve exhibits many interesting 
features. The first notable feature was a sluggish rise to 
visual maximum.  After discovery, V1186 Sco required 6 days 
to reach its maximum at V = 9.6.  Very few novae have exhibited such long 
pre-maximum behavior which is generally only found in very slow 
novae. 
Another intriguing feature of the V1186 Sco light curve 
is the strength of the second maximum that occurred about 9 days later after
the first.  This second maximum was extremely bright, only $\sim$ 0.2 
mag fainter than the primary maximum.  If not for the $\sim$ 1 mag decline 
twelve days after discovery, the early light curve evolution might be better
described as a plateau, similar to that observed in V723~Cas.  While secondary 
maxima are not unusual in slower nova they do not usually approach that of 
the primary maximum \citep[see ][for numerous examples of nova light 
curves]{PG57}.  For example, PW~Vul displayed multiple 
secondary peaks in its light curve but they were all more than a magnitude 
fainter than at maximum.  

After the second peak in the V1186 Sco light curve, the V magnitude showed 
a rapid decline passing its t$_2$ point (the time to decline 2 
magnitudes from maximum) in 20 days.  Subsequently, the V1186 Sco light 
curve had two smaller peaks during its decline.  These later peaks 
make it difficult to establish the t$_3$ time (the time to 
decline 3 magnitudes from maximum) since the brightness fluctuated around 
V = 12.6.  V1186 Sco first reached V = 12.6 after 45 days but then 
brightened and subsequently did not become fainter than V = 12.6 until 
70 days after maximum.  Which decline value is representative of 
this nova?  The choice has significant consequences 
on speed-class classification \citep{Warner95}, WD progenitor 
identification, and any physical parameters based on the derived decline 
times.  After 100 days the V band light curve stabilized at about 13.5 
magnitudes as the fading continuum was offset by emerging line emission, 
particularly from [\ion{O}{3}] (5007\AA).  According to Duerbeck's study 
of about 100 galactic novae \citep{Duerbeck81}, the light curve of V1186 Sco
is classified as type Bb (``decline with major fluctuations, {\it e.g.,} 
double or multiple maxima'').  The prototypes of this classification are 
\object{DN Gem} and \object{NQ Vul} \citep[see Fig. 3 in ][]{Duerbeck81}.
There was no evidence for a major visual extinction event characteristic 
of dust formation in the visible light curve of V1186 Sco during the 
first $\approx 300$ days. Dust is not frequently found in novae with
short $t_{2}$ timescales, V1974~Cyg or V838~Her \citep{Woodward97,Woodward92} 
being recent examples.

\subsection{SPECTROSCOPY}

Multiwavelength spectroscopy, centered around the \textit{Spitzer} 
observations, was obtained during two different epochs.  The first, called 
E1, occurred $\sim$60 days after visual maximum between 22 August 2004 and
23 September 2004.  This was the earliest that \textit{Spitzer} 
data could be obtained after
the ToO trigger was requested.  The next set of \textit{Spitzer} observations 
could not be scheduled and executed until the following year due to object 
visibility window constraints.  This second epoch, called E2, was taken 
during the transition to the nebular phase, 21 May 2005 - 16 July 2005, 
or over 261 days after maximum.  Table~\ref{tab:spec-obslog} provides the 
relevant observational information for all the spectral observations while
Tables~\ref{tab:fepoch1} and \ref{tab:fepoch2} give the measured line
fluxes, uncorrected for reddening, obtained during the E1 (2004;
\textit{Spitzer} only) and E2 (2005; all data) epochs, respectively.

\subsubsection{Spitzer mid-IR}


V1186 Sco was observed by the \textit{Spitzer Space Telescope} 
\citep{Gehrz07a,W04} four times during the two observational epochs 
using the Infrared Spectrograph \citep[(IRS)]{H04}. The observations 
were part of the Cycle 1 classical nova ToO initiative, program 
identification number (PID) 2333 (PI C.\ E.\ Woodward). 

Basic Calibrated Data (BCD) products were calibrated and processed with 
the {\it{Spitzer}} Science Center (SSC) data pipeline.  The IRS 
observations were conducted using the short wavelength (5.2 - 14.5~\micron)  
low resolution module (SL), the short wavelength (9.9 - 19.5~\micron) 
high resolution module (SH), and the long  wavelength (18.7 - 37.2~\micron) 
high resolution module (LH).   The resolving power for SL is $R$ = 
$\lambda/\Delta\lambda \sim 60-128$, while R $\sim$ 600 for SH and LH. 
Emission lines observed with the latter modules are marginally resolved. 

All \textit{Spitzer} observations utilized visual PCRS peak-ups on nearby 
isolated stars to ensure proper placement of the target in the narrow IRS 
slits. The spectroscopy of V1186 Sco during E1 (AORs 10273536, 10273792, 
10274048) consisted of 2 cycles of 6 second ramps in SL (12 
seconds on-source), and 6 cycles of 6 second ramps in SH and LH (36 seconds 
on-source each), while the E2 observations (AORs 10274304, 10275328) 
employed 2 cycles of 30 second ramps (60 seconds on-source) for SH, 3 
cycles of 60 second ramps for LH (180 seconds on-source), and 3 cycles of 
6 second ramps (18 seconds on-source) for SL.  IRS basic calibrated 
data products (BCDs) were processed with version 12.0.2 of the IRS 
pipeline. Descriptions of the IRS instrument and its operation are 
available in \cite{H04} and \cite{Gehrz07a}. Details of the 
calibration and raw data 
processing are specified in the IRS Pipeline Description Document, 
v1.0.\footnote{http://ssc.spitzer.caltech.edu/irs/dh/PDD.pdf}. 
E1 epoch observations were obtained with a short synoptic cadence 
($1-2$~day), and no marked evolution in the spectral behavior 
of V1186~Sco was noted within the uncertainty of the absolute IRS
calibration errors. Thus, the three separate spectra were averaged 
together to produce a single composite spectrum for analysis. 

Bad pixels were interpolated in individual BCDs using bad pixel masks 
provided by the SSC. Multiple data collection events were obtained at two 
different positions on the slit using \textit{Spitzer's} nod 
functionality.  Sky subtraction was only possible for the SL observations, 
as no dedicated sky observations were performed for the SH and the LH mode 
observations. Sky subtraction was performed by differencing the two 
dimensional SL BCDs to remove the background flux contribution.  Spectra 
were then extracted from the background corrected SL data and the SH 
and LH BCDs with SPICE (version 1.3-beta1) using the default point source 
extraction widths. The extracted spectra were then combined using a 
weighted linear mean into a single output data file.  At the time of 
reduction, the errors generated by the SSC pipeline were not reliable and 
therefore errors were estimated from the standard deviation of the flux at 
each wavelength bin. The spectral lines detected in the \textit{Spitzer} 
data were fit using a non-linear least squares Gaussian routine \citep[the 
Marquardt method,][]{Bev92} that returns the line center, line amplitude, 
integrated line flux, continuum amplitude and the slope of the continuum.  


\subsubsection{Ground-based Near-IR}

Three telescopes were used to obtain the NIR spectral data.  The 
first observation was obtained at the Keck II telescope using NIRSPEC 
\citep{Mc98}.  All of the observations were in low-resolution mode with 
the 42\arcsec \ by 0.38\arcsec \  slit.  The seeing during the observation 
was about 0.8\arcsec.  Multiple spectra using different sorting filters 
were taken by nodding V1186 Sco and standards (HD 162220 and HD 163633) 
along the slit at 15\arcsec \ intervals while guiding.  The REDSPEC 
software was used to reduce the nova and standards using standard IR 
techniques \citep{Joyce92}.  The N-1 filter spectrum was flux calibrated 
using the standard's VJHK magnitudes normalized to the 1 \micron\ point of a 
T$_{eff}$ = 9480~K blackbody.  The other filters were calibrated using the 
2MASS magnitudes of the standards. The resulting fluxes are only 
approximate because of slit losses. 

Additional medium-resolution NIR spectroscopy was obtained from the United 
Kingdom Infrared Telescope (UKIRT)\footnote{The United Kingdom Infrared
Telescope is operated by the Joint Astronomy Centre of behalf of the United
Kingdom Particle Physics and Astronomy Research Council.} using the 
facility spectrographs CGS4 \citep{Mountain90} and UIST \citep{Ramsay04} 
to cover 0.8--1.35~$\mu$m and 1.4--5.1~$\mu$m bands, respectively, at 
resolving powers of 500--2000. The spectra were obtained through 
UIST's 0.24\arcsec \ slit and CGS4's 0.6\arcsec \ slit in 
the conventional stare/nod-along-slit mode. The data were reduced using
standard Figaro procedures, utilizing contemporaneous arc lamp spectra
for wavelength calibration and spectra of nearby early type bright stars
for flux calibration and removal of telluric absorption lines.

The final NIR data set was acquired from the Infrared Telescope Facility (IRTF)
using SpeX \citep{Rayner03}.  The data were obtained through 0.8\arcsec \ by 
15\arcsec \ slits and a 10\arcsec \ N-S nod for background cancellation.  
Due to the proximity of calibrators (\object{HD 161822} and 
\object{HD 157486}), chopping 
and extinction corrections were not necessary.  SpeXTools \citep{C04} was 
used to reduce the data. Calibrator V band magnitudes and spectral types 
from the Bright Star Catalog \citep{HJ82} and matched 
\citet{Kurucz91,Kurucz94} spectral energy distribution (SED) models were 
used to create the flux models.  The absolute flux scale was set by 
normalizing to the calibrators' K band magnitudes.  These were calculated 
from the known V band magnitudes and (V-K) colors of stars of similar 
spectral type \citep{Koor83} as the calibrators. 

\subsubsection{Optical}


The first optical spectrum was obtained at the Multiple Mirror Telescope 
with the Blue Channel Spectrograph \citep{SWF89}.  The slit was in 
long-slit mode (1\arcsec \ by 180\arcsec) with a 300 line/mm grating 
centered at 6000\AA \ giving a resolving power of $\sim$ 900. The second 
optical spectrum was taken at Steward Observatory's Bok
Telescope with the B\&C spectrograph. Observations
were obtained using a 1.5\arcsec \ by 240\arcsec \ slit and the
400 line/mm grating with a central wavelength of 5038\AA. The standard 
stars used for the calibration were Feige 34 and HZ 44. All 
optical data were reduced using standard IRAF packages\footnote{IRAF is 
distributed by the National Optical Astronomy Observatories, which are 
operated by the Association of Universities for Research in Astronomy, 
Inc., under cooperative agreement with the National Science Foundation.} 
and spectral extraction techniques. 

\section{REDDENING AND DISTANCE}
\label{sec:redist}

An accurate determination of the reddening toward a nova is of critical 
importance for obtaining corrected line ratios and determining other basic 
parameters such as the distance.  Optical spectra obtained by M.~Fujii 
near visual maximum clearly showed \ion{Na}{1} D absorption lines indicative
of moderate reddening \citep{Fujii04}.  The \ion{Na}{1} D absorption 
line is present in our earliest optical spectrum but because of the
complexity of the blended lines a determination of the reddening is not
possible.  Thus, other indirect methods must be used to determine the 
reddening toward V1186 Sco.  

\citet{VY87} suggest 
two ways to estimate the reddening of a nova using the 
early light curve decline and colors.  They report that at maximum and at 
t$_2$ the average intrinsic (B-V) colors for novae are 0.23$\pm$0.06 and 
$-0.02\pm0.04$, respectively.  Using the observed (B-V) colors for V1186 
Sco (see the lower portion of Fig.~\ref{aavso-bb}b) the E(B-V) 
values are consistent
at 0.45$\pm$0.1 and 0.47$\pm$0.16, respectively.  However, the value of 
the line of sight extinction toward the nova inferred from the reddening 
map of \citet{schlegel98} yield a larger value of E(B-V) = 0.96.  Unfortunately,
both methods have serious limitations.  As described in \S2.1, the 
light curve of V1186 Sco does not exhibit the smooth decline of a typical 
nova.  Instead the light curve has numerous secondary peaks which make a 
determination of the true decline times problematic while the 
\citet{schlegel98} extinction map is not as reliable at the location
of V1186 Sco, {\it e.g.} when $b \le$ $|5|\arcdeg$.

Alternatively one can infer the reddening by comparing the observed 
emission line ratios to those predicted under optically thin conditions.  
For hydrogen lines we use the predicted hydrogen line ratios from 
\citet{HS87} at T$_e$ = 10$^4$ K and N$_e$ = 10$^8$ cm$^{-3}$.  This is a 
typical temperature for a nebular gas and the density is consistent with 
an ejected mass $\sim$ 10$^{-4}$ M$_{\odot}$ assuming ejecta traveling in 
a spherical shell with inner and outer velocities of 500 and 
1000~km~s$^{-1}$.  We use the Pa$\gamma$/Br$\gamma$ ratio in the NIR, as the
Balmer decrement may not yet be radiative.  The observed Pa$\gamma$/Br$\gamma$
was 2.6, corresponding to E(B-V) value of 0.43. Another line ratio method 
uses the ratios of the NIR \ion{O}{1} 0.8448 and 1.128 $\mu$m lines, with 
a correction from the 1.316 $\mu$m line, to determine reddening \citep{rudy91}.
This method produces an E(B-V) = 0.45 using the last NIR data set.  With the 
exception of the \citet{schlegel98} extinction map, all the reddening methods 
converge toward E(B-V) = 0.45$\pm$0.1 which we adopt as the reddening for
V1186 Sco.

Most distance estimates for novae depend on the relationship between the
maximum magnitude versus the rate of decay of the light curve (MMRD).  
Using the MMRD of \citet{DVL95} and t$_2$ = 20 days provides an 
M$^{max}_V = -8$.  With the adopted reddening
the derived distance is 17.3 kpc or well beyond the Galaxy. It is highly 
unlikely that V1186 Sco is this distant.  Another approach is to assume 
the same absolute magnitude of a nova at a known distance and with similar 
properties.  PW Vul is a good candidate with a comparable light curve and 
emission line evolution.  An expansion parallax measurement for PW Vul 
\citep{DD00} gives M$^{max}_V = -6.7$.  Adopting this absolute magnitude 
places the V1186 Sco distance at 9.5$^{+1.4}_{-1.2}$ kpc.  With V1186 Sco's 
position toward the Galactic center, even this distance puts V1186 Sco on 
the other side of the Galaxy and is incongruous with the moderate reddening
derived above.  As we will show in \S\ref{sec:phtmd}, 
photoionization modeling requires a significantly lower M$_V$, corresponding 
to a closer distance of 5.5~kpc.  Given the erratic nature of the light 
curve and its location near the Galactic center, we adopt this lower 
distance as the most realistic distance to V1186 Sco.

\section{LINE EVOLUTION}
\label{sec:linevol}

\begin{figure}
\plotone{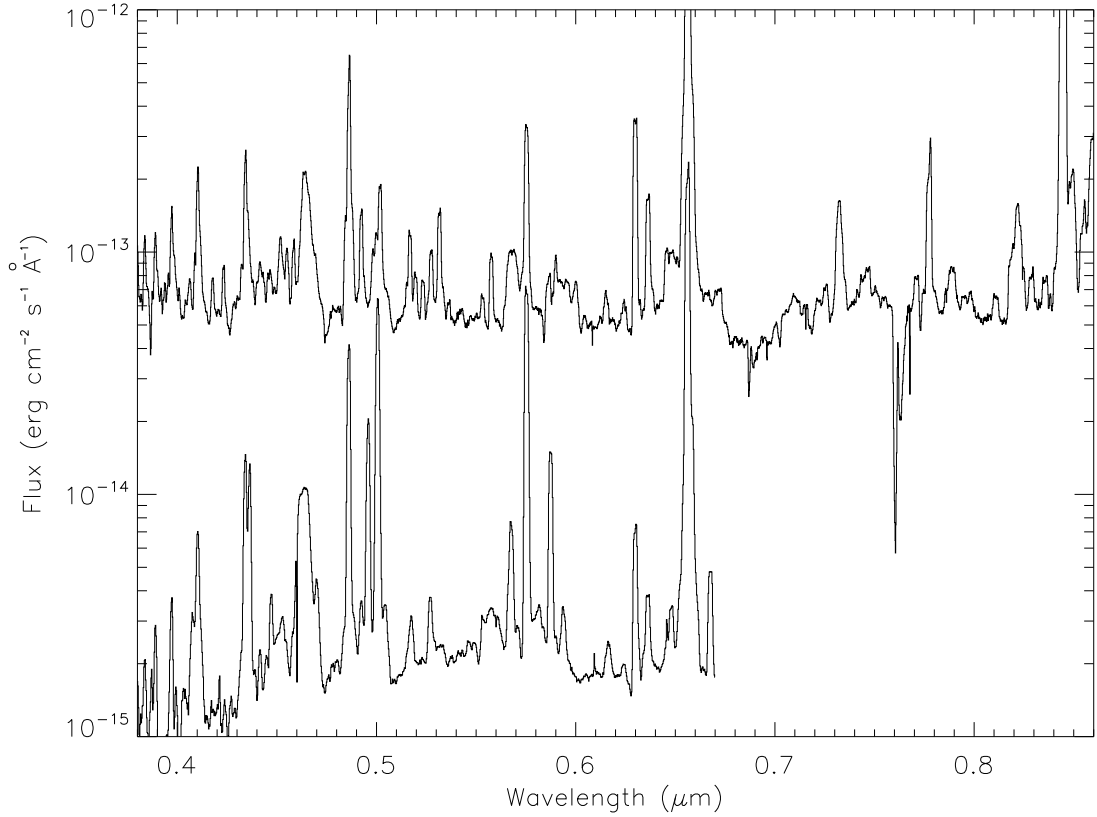}
\caption{The evolution of the optical spectrum during the two observational
epochs.  The first spectrum ({\it top} line) is typical of a \ion{Fe}{2}
type novae during the early outburst with narrow, low ionization emission 
and numerous P Cygni profiles.  The second spectrum ({\it bottom} line) shows 
an emerging nebular spectrum.
\label{optical-spec}}
\end{figure}

\begin{figure}
\plotone{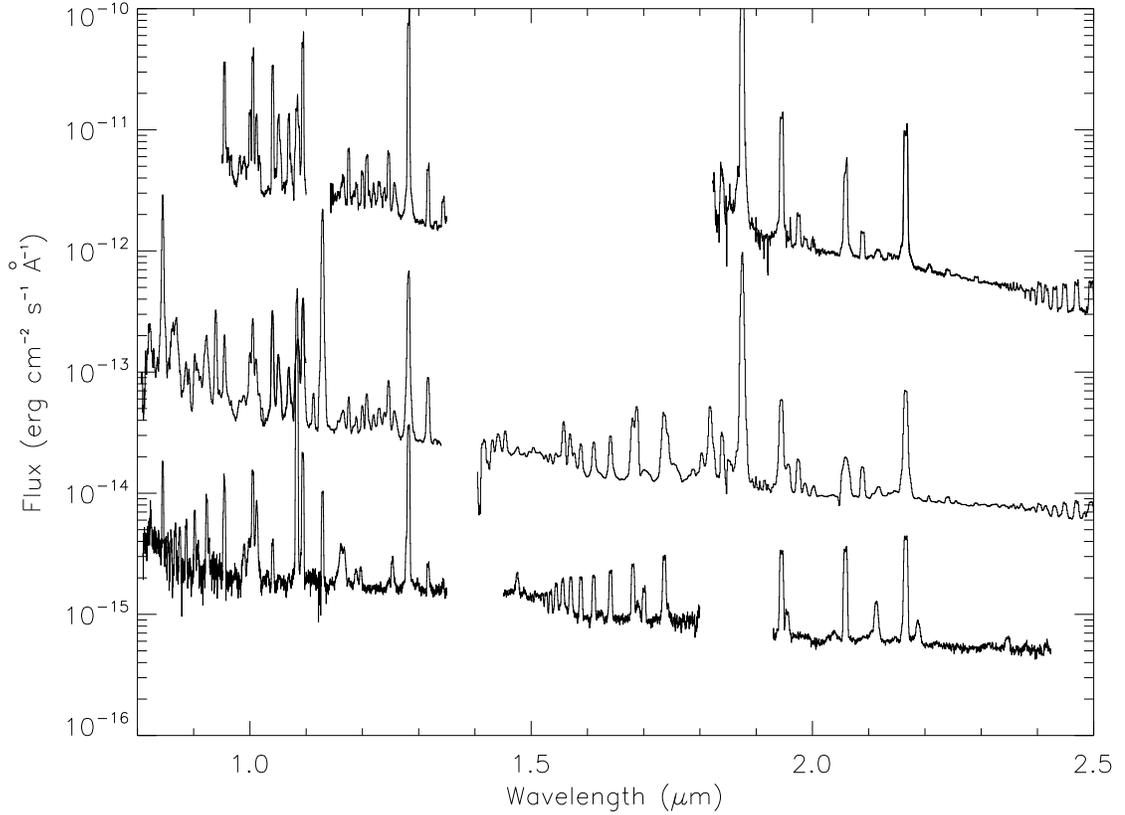}
\caption{The evolution of the near-IR spectrum. The Keck ({\it top} line) and
UKIRT data ({\it middle} line) were obtained early in the outburst in the
first epoch.  The last SpeX spectrum ({\it bottom} line) was taken over a year
after outburst.  The strongest emission is from \ion{He}{1} and \ion{O}{1}.
The first SpeX spectrum is not displayed because it is
of similar signal-to-noise as the {\it middle} UKIRT data obtained 1-2
nights later, and exhibiting the same suite of emission lines
in the {\it JHK}-bands.
\label{a_nir-spec}}
\end{figure}

\begin{figure}
\plotone{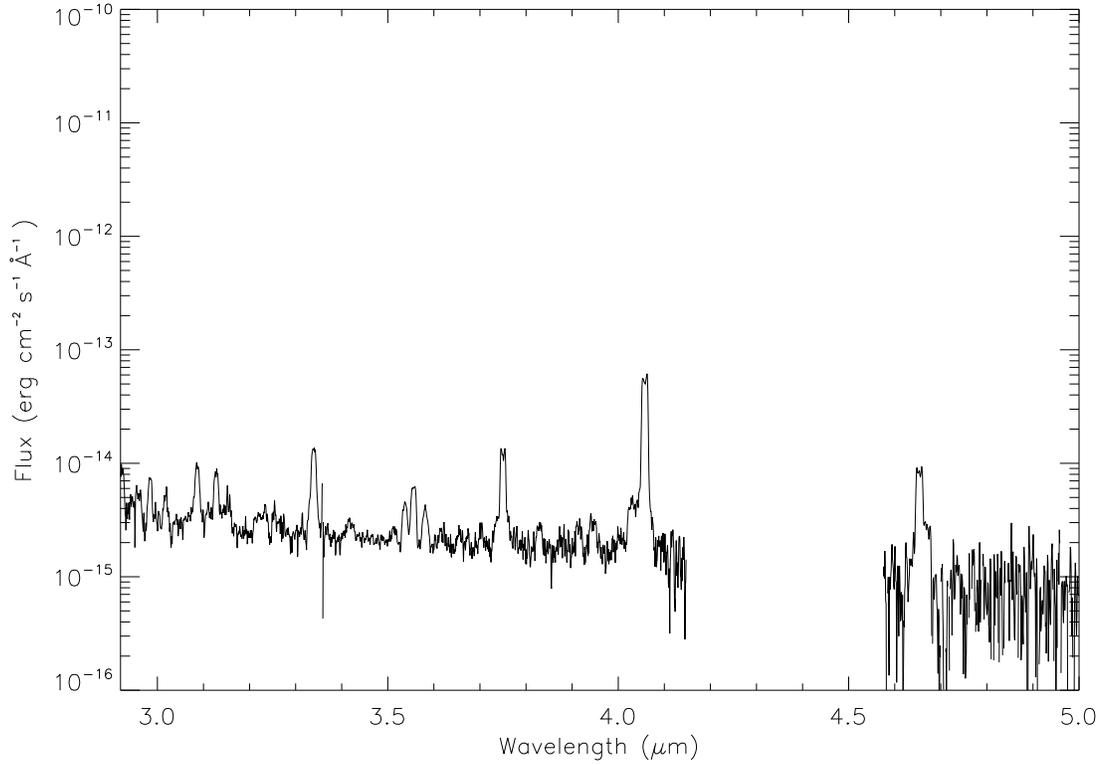}
\caption{The {\it LM}-band spectra obtained at UKIRT. The spectrum
at this epoch (03 September 2004) is dominated by hydrogen recombination 
lines, $[$\ion{N}{1}$]$, little continuum, and no 3~\micron \ dust emission 
features. The {\it M}-band spectra is truncated near 5.0~\micron \ due to 
poor signal-to-noise.
\label{b_nir-spec}}
\end{figure}

\begin{figure}
\plottwo{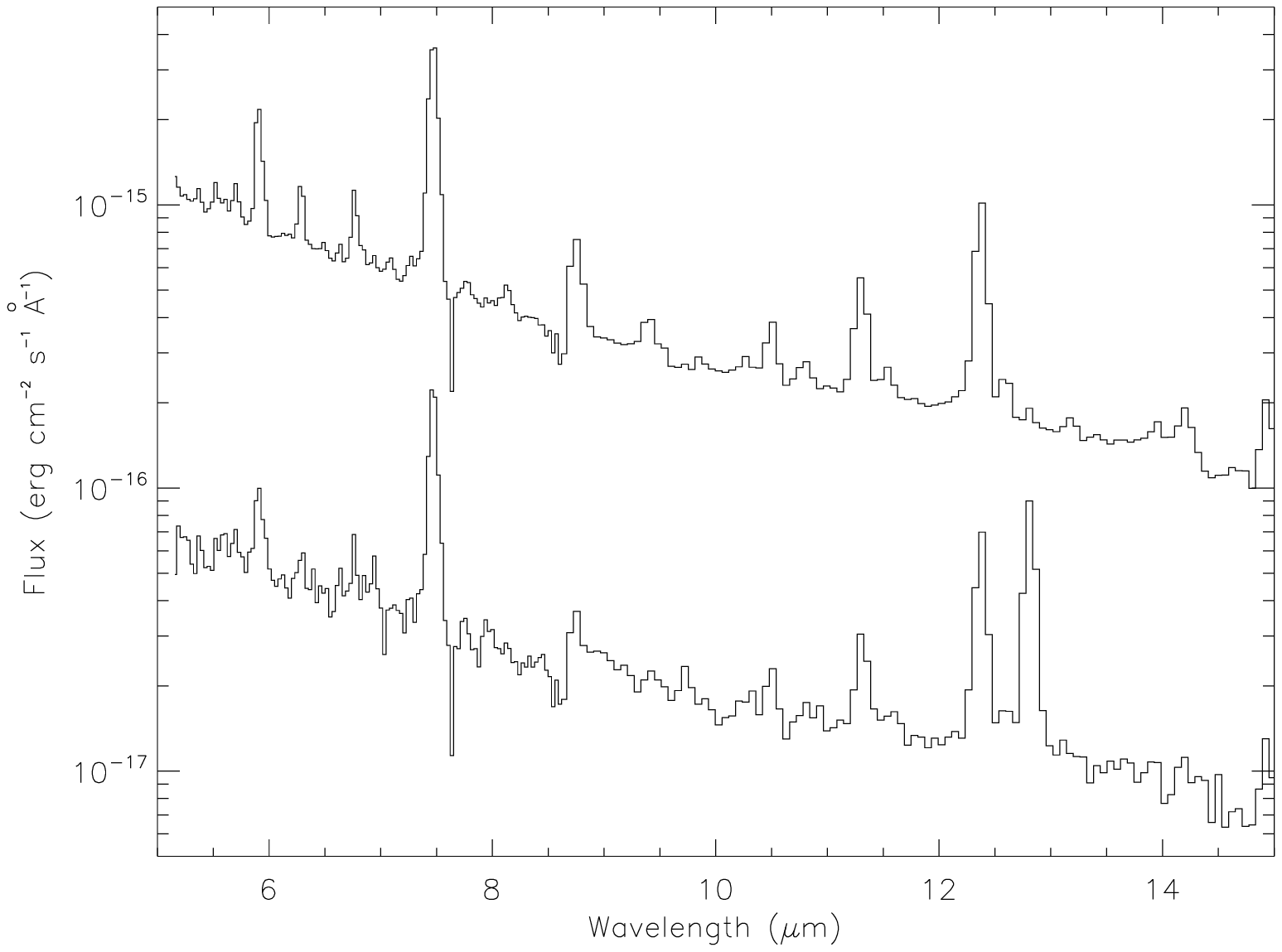}{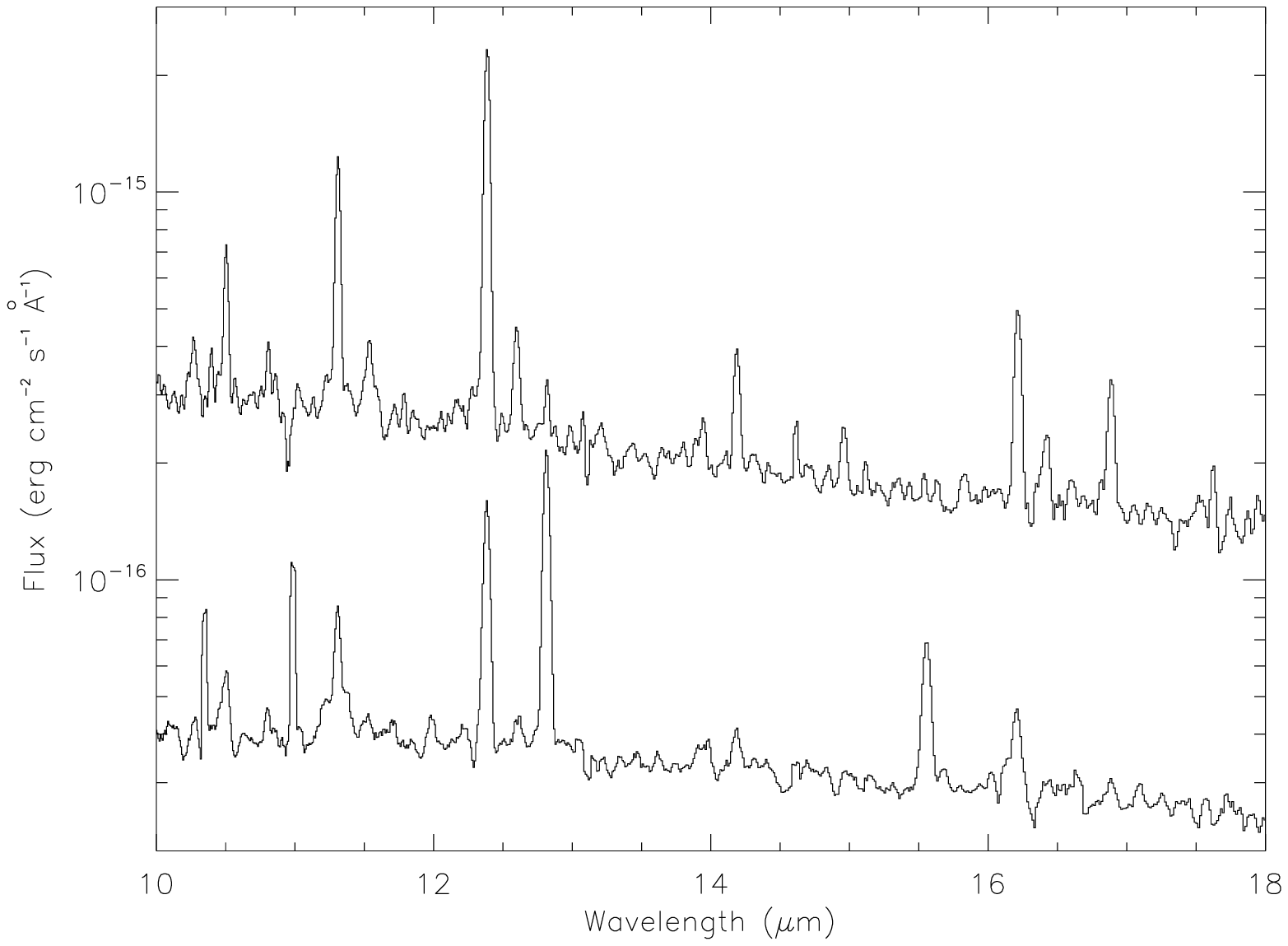}
\caption{The evolution of the \textit{Spitzer} SL ({\it left} panel) and SH 
({\it right} panel) spectra from the two observational epochs.  The first 
epoch ({\it top} line) is dominated by a strong hydrogen recombination 
spectrum.  Weak [\ion{Ne}{2}] (12.81 $\mu$m) can be seen in the SH
spectrum.  The hydrogen lines are still present at the second epoch 
({\it bottom} line) but the IR fine-structure lines ([\ion{Ne}{2}] and 
[\ion{Ne}{3}] at 15.56 $\mu$m) are now the strongest lines 
present. The later spectra are offset by 0.3 Jy for clarity.
\label{spitzer-spec}}
\end{figure}

The line evolution from E1 (2004) to E2 (2005) in the optical, NIR, and
far-IR is shown in Figs.~\ref{optical-spec} through \ref{spitzer-spec}. 
The E1 data are similar in 
structure to other CO type novae at the t$_3$ time.  In the optical 
V1186 Sco was transitioning from the optically thick to an optically thin 
phase.  The \ion{Fe}{2} emission typical of early CO nova had faded and 
the only forbidden lines seen were [\ion{N}{2}] and [\ion{O}{1}].  In the 
NIR, the spectra were dominated by hydrogen recombination and emission from 
neutral oxygen and helium. The E1 \textit{Spitzer} spectra showed strong 
hydrogen recombination lines and surprisingly, a weak [\ion{Ne}{2}] 
(12.81$\mu$m) line.  The emergence of this line is typically taken as a sign 
that the outburst occurred on the surface of an ONeMg WD \citep{Gehrzet98}
although \citet{greenhouse88} and \citet{Gehrzet88} and others have argued 
that strong continuum emission from condensing dust in the ejecta may mask
the presence of [\ion{Ne}{2}] lines, especially if observations are 
obtained at low spectral resolution.  Previous [\ion{Ne}{2}] detections
discussed in the literature, \object{V1974 Cyg} \citep{hayward92}, 
\object{V1187 Sco} \citep{lynch2006}, \object{QU Vul} 
\citep{gehrz07b,greenhouse88,ggh85}, and \object{CP Cru} \citep{lyke03}, 
occurred on ONeMg nova with high ejection velocities ($\ge 1000$~km~s$^{-1}$).
However, it is unlikely that the progenitor is a ONeMg WD in the V1186 Sco 
system as the speed class (slow) and expansion velocities from the line 
widths are more similar to that observed in CO WD outbursts.

During the next epoch all the spectra are different but the hydrogen 
lines are still prominent.  Almost a year after outburst the nova was just 
entering the nebular phase with the emergence of [\ion{O}{3}] in the 
optical.  The [\ion{N}{2}] and [\ion{O}{1}] lines observed earlier were 
still present but \ion{He}{1} lines also were present.  The overall 
ionization of the ejecta at this point was still low with no typical high 
ionization lines ({\it e.g.,} [\ion{Fe}{7}] 6087) present in the optical 
spectrum.  In the E2 NIR spectrum the emission lines had narrowed 
due to the expansion of the ejecta and 
neutral oxygen was still present even one year after outburst.  The 
strongest line was the \ion{He}{1} triplet at 1.08 $\mu$m which had 
increased in intensity by more than a factor of 30.  A similar increase 
was observed but earlier, between 130 and 230 days after maximum in PW Vul 
\citep{Will96}.  That the line evolution of V1186 Sco was slower
than observed in PW Vul, strengthens our earlier contention that V1186 Sco's 
M$_V$ value must also be fainter than PW Vul's.

By E2, the [\ion{Ne}{2}] emission observed earlier was now prominent 
in the \textit{Spitzer} spectra along with [\ion{Ne}{3}] (15.56 $\mu$m).
The poor S\/N of our E2 optical spectrum below 0.4 \micron\ makes it 
impossible to determine if there was any corresponding strong [\ion{Ne}{3}].
The 18 to 34 \micron\ spectrum was very noisy but showed clear detections of 
\ion{H}{1} (9-8, 12-10, 13-10, and 15-11).  Other than the hydrogen and
neon lines, no other lines were detected in the \textit{Spitzer} spectra.

\begin{figure}
\plottwo{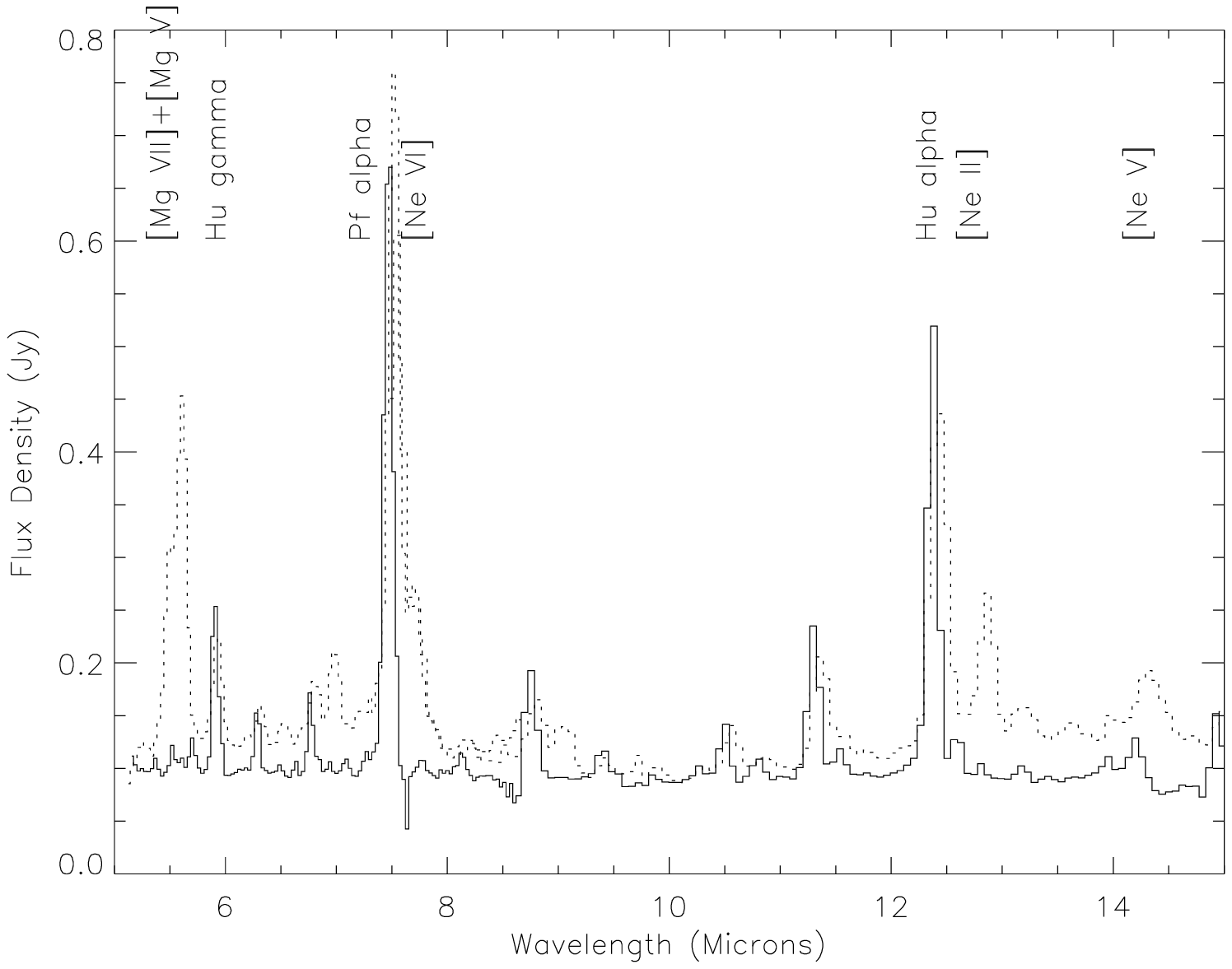}{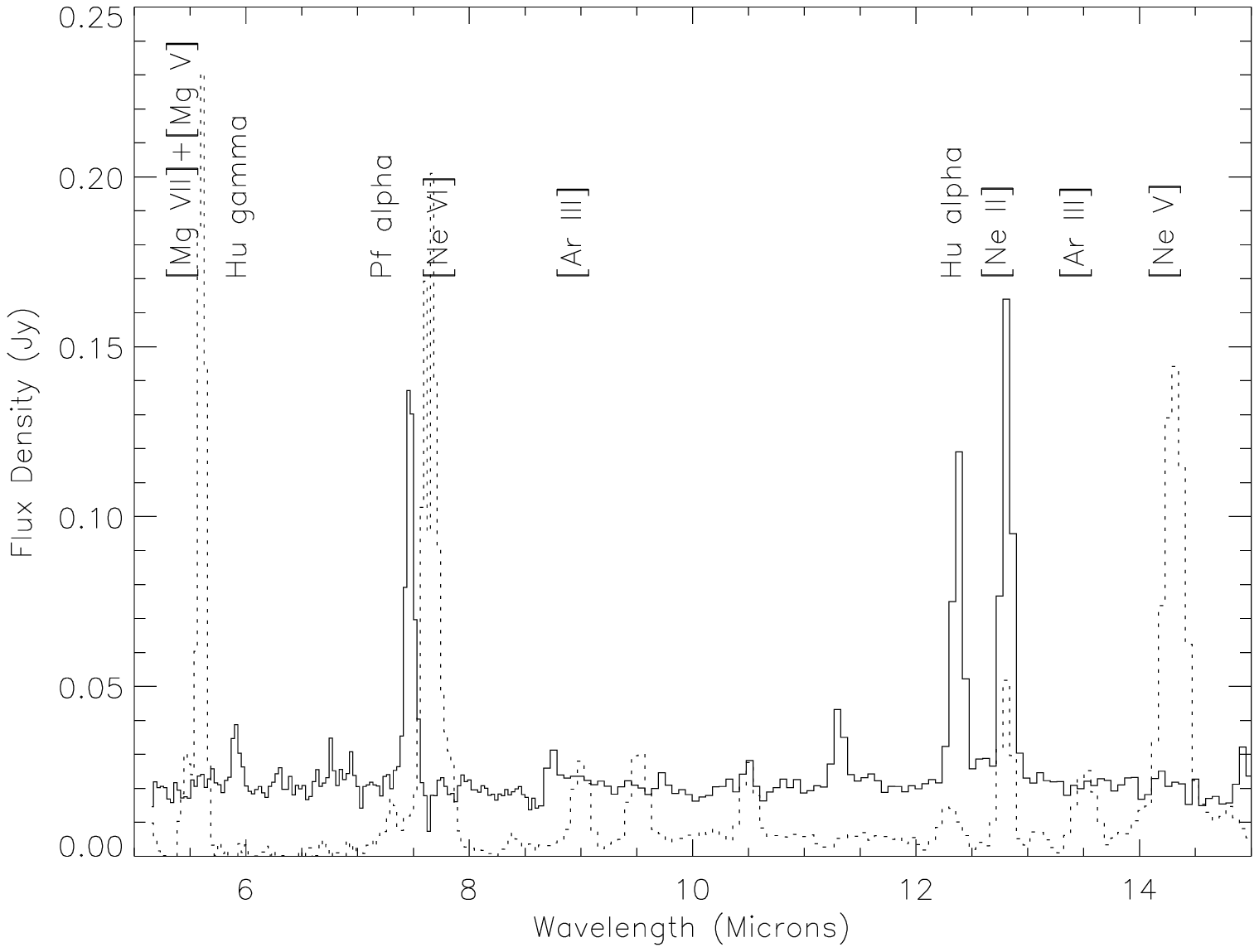}
\caption{Comparison of \textit{Spitzer} IRS SL line emission evolution 
in nova V1186 Sco ({\it solid} line) and the much faster, ONeMg 
nova V1187 Sco ({\it dashed} line).  The {\it left} panel compares
the 29 August 2004 spectrum of V1186 Sco (E1), with the 28 September 2004~UT
spectrum of V1187 Sco observations, while the {\it right} panel compares
the 21 March 2005~UT (E2) of both novae obtained on the same day. V1186 Sco,
in contrast to V1187 Sco, did not develop strong emission lines from
metals, such as Mg or Ar, that are commonly detected in ONeMg
novae.
\label{v87_86-spec}}
\end{figure}

To emphasize the slow evolution in V1186 Sco, Fig.~\ref{v87_86-spec} 
compares the \textit{Spitzer} observations of V1186 Sco with the faster 
ONeMg nova V1187 Sco \citep{lynch2006}.  V1187 Sco reached visual maximum on 
03 August 2003~UT, or approximately one month after V1186 Sco.  With t$_2 
\simeq 9$~days and t$_3 = 15$~days, its decay rate was significantly faster
than V1186 Sco. \textit{Spitzer} spectra of both novae in Scorpius were 
obtained within a few days of each other. The first epoch \textit{Spitzer} 
spectra of V1187 Sco (see Fig.~\ref{v87_86-spec}a) already displayed 
[\ion{Ne}{2}] in addition 
to higher ionization lines such as [\ion{Ne}{6}] (7.65 \micron) and 
[\ion{Ne}{5}] (14.3 \micron).  In the second epoch observations of V1187 Sco 
(Fig.~\ref{v87_86-spec}b) the hydrogen emission is almost gone and 
the higher ionization state lines continued to increase in 
strength.  In addition to the prominent neon 
lines, V1187 Sco had strong magnesium and argon lines typical of a fast 
ONeMg nova. 

\subsection{PHOTOIONIZATION MODELING}
\label{sec:phtmd}

The available data set poses challenges for photoionization modeling. Due 
to its lethargic evolution, V1186 Sco was only entering the nebular phase 
in the E2 (2005) epoch.  The majority of the dominant lines are hydrogen 
recombination lines with relatively few forbidden lines available for 
analysis.  Another dilemma is that there currently are no additional 
published data sets obtained at later epochs (post-2005) to verify any 
model solution of the E2 observations.  Finally, the E2 data set 
was obtained over a 100 day span.  However, despite these issues it is possible 
to determine an abundance solution by 1) assuming that the conditions 
in the ejecta evolved slowly enough that a single model is appropriate for 
all the spectra obtained during the E2 epoch and  2) fitting the 
emission lines from each spectral region relative to itself ({\it e.g.,} 
optical to H$\beta$, NIR to Pa$\beta$, and the \textit{Spitzer} data 
to Hu$\alpha$) to account for uncertainties in the absolute flux 
calibrations and the different times the data were obtained.

We used the \cldy\ 2006b photoionization code \citep{Fer98} and assumed a 
spherical shell ejection with inner and outer radii set by material moving at 
500 and 1000 km s$^{-1}$ over 260 days.  The density profiles assumes a 
r$^{-3}$ profile to provide a constant mass per unit volume throughout the 
model shell. The model parameters that were allowed to vary were the effective 
temperature and luminosity of the source (assumed to be a blackbody), and
the inner hydrogen density.  The filling factor was set at 0.2, typical
of what we have found in previous photoionization models of classical novae 
\citep[see][and references within]{Schwarz07}.  The abundances of elements
with observed emission lines (helium, nitrogen, oxygen and neon) were 
also allowed to vary while all other elements were left at 
their solar abundances.  The model was tuned using the 
$[$\ion{Ne}{2}$]$ (12.81\micron)/$[$\ion{Ne}{3}$]$ (15.56\micron) 
ratio which was sensitive to the effective temperature of the photoionizing 
source while the $[$\ion{O}{3}$]$ (0.4363\micron)/$[$\ion{O}{3}$]$ 
(0.5007\micron) ratio constrain the electron density.  The results 
of the best fit model to the 35 observed line ratios are presented in 
Table~\ref{tab:suitlines} and the best fit model parameters are given 
in Table~\ref{tab:cldyparam}.  With seven free parameters, the best fit
model had $\chi^2$ = 38, or a reduced $\chi^2$ of 1.36.

The best fit model had a luminosity of 6.3$\times$10$^{36}$ erg s$^{-1}$ and 
was extremely cool, T$_{eff}$ = 47,000 K.  This low model 
temperature confirms that the ejecta of V1186 Sco was evolving even more 
slowly than PW Vul.  The infrared spectra of PW Vul taken obtained 270
days after outburst had already revealed 
the emergence of coronal lines of 
[\ion{Si}{6}] (1.961\micron) and [\ion{Mg}{8}] (3.026\micron)
\citep{Will96}.  There are none of these high ionization coronal lines
in the NIR spectrum of V1186 Sco more than a year after outburst. 
The ejected mass of the best V1186 Sco 
model was 6.5$\times$10$^{-5}$ M$_{\odot}$.  This mass is within a
factor of 2 of the typical ejected mass of $\sim$ 10$^{-4}$ M$_{\odot}$
found for most novae \citep{Schwarz97,Schwarz01,Schwarz07} using similiar
photoionization modeling techniques.

The helium, nitrogen, oxygen, and neon abundances were all enhanced 
at 1.3, 50, 4.7, and 1.3 times solar, respectively. In addition, the 
best fit model with sulfur, argon and iron at solar abundances also predicted 
emission lines that were not observed, notably [\ion{S}{3}] (0.9069$\mu$m),
[\ion{Ar}{3}] (9.0$\mu$m), and \ion{Fe}{2} (0.62$\mu$m).  Reducing these
abundances by a factor of 2 removed the sulfur and argon lines 
and greatly reduced the predicted \ion{Fe}{2} and \ion{Fe}{3} lines.  

While the oxygen and nitrogen abundances are typical of CO type novae 
\citep{Schwarz97,Schwarz01}, the derived neon abundance is interesting.  
The slight enhancement is not as much as would be expected for a ONeMg type 
nova \citep[Ne $\gtsimeq$ 20, see][]{Schwarz07} but more than expected for 
an outburst occurring on a pure CO white dwarf.  In that case the 
neon abundance should be similar that of the other heavier elements 
such as sulfur, argon, and iron 
which reflect the composition of the secondary.  To determine if the 
derived neon enhancement is significant we estimated the uncertainty in the 
model abundances.  Ideally, we would use multiwavelength observations 
taken during a more evolved epoch to see how well the 
abundance solution fit.  Unfortunately, there are no later data 
available thus we used the method outlined in
\citet{Schwarz01} to estimate the uncertainties in the abundances. 
The technique assumes that the model parameters were not correlated.
Holding all other parameters fixed at their values in 
the best fit model, one abundance at a time was adjusted until 
the reduced $\chi^2$ of the model increased to 2.  This approach provides 
an approximate 3$\sigma$ uncertainty for each abundance.
The neon abundance range from this method is
0.7 - 1.9 times solar (see Table~\ref{tab:cldyparam}) which is still greater 
than the upper limit of 0.5 times solar for the sulfur, argon, and 
iron abundances.  Note that the derived neon abundance is relative to the
solar neon value of \citet{GN93} which is similar to the new results of 
\citet{Cunha06}.  Both neon abundances are significantly higher, approximately
0.3 dex, than the recent solar neon values of \citet{Lodders03} and 
\citet{Asplund05}.  If the solar neon abundance is closer to that determined
by \citet{Lodders03} and \citet{Asplund05} then our derived neon abundance
is even more impressive.

\begin{figure}
\plotone{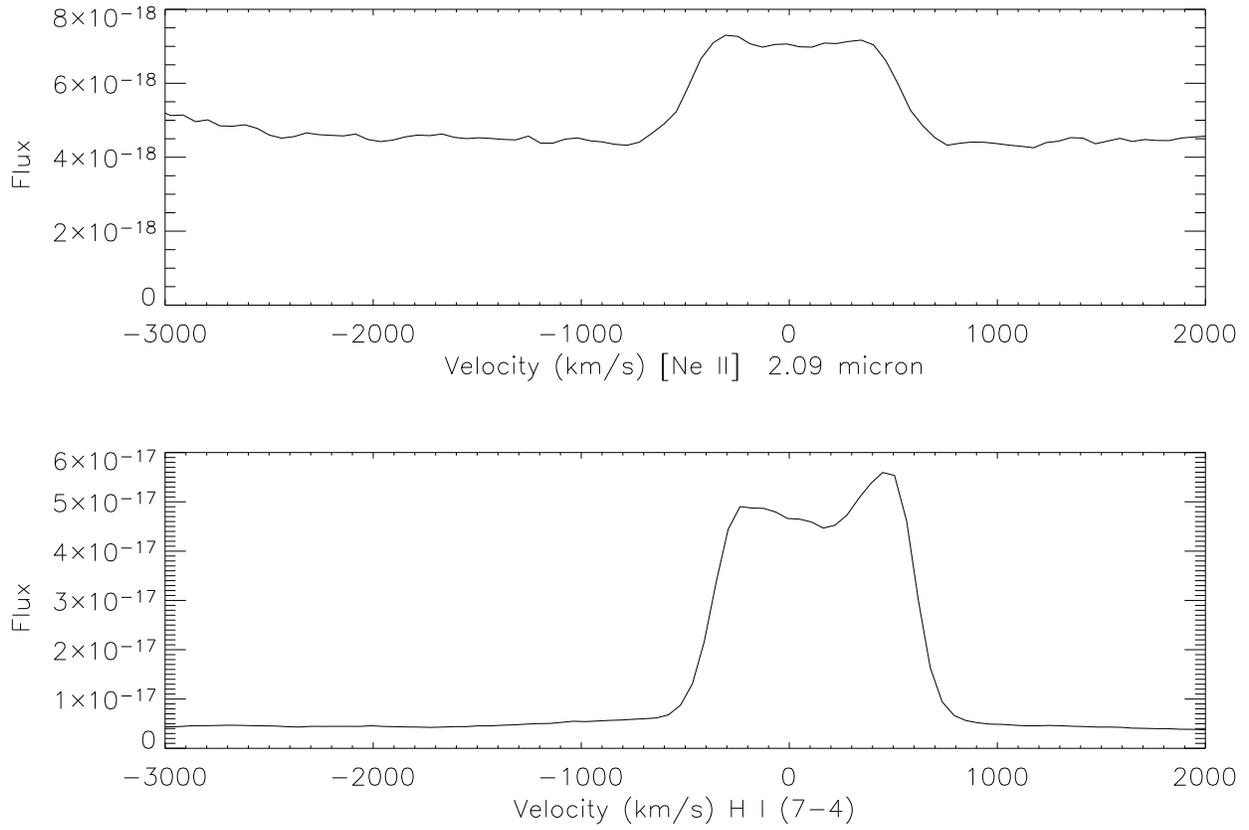}
\caption{Early line profiles of [\ion{Ne}{2}] (2.09$\mu$m) and \ion{H}{1} (7-4)
showing the asymmetry between the forbidden lines and the hydrogen lines.
\label{profiles}}
\end{figure}

The high resolution line profiles from the Keck observations indicate 
that the \ion{H}{1} profiles (all series from n=2, 3, and 4) have the 
same asymmetry throughout the observed period.  The identical nature of 
all these transitions implies that the optical depth is likely small 
enough to insure the correctness of our extinction inferred in Section 3.
The \ion{H}{1} profile similarities is confirmed by the asymmetries 
evident in the \textit{Spitzer} high resolution spectra although the 
profiles are not well resolved.  In contrast, the 
forbidden lines, for instance [\ion{Ne}{2}] 2.089$\mu$m, display 
nearly symmetric profiles (Fig.~\ref{profiles}). The shell thus 
appears, as in many novae observed with sufficient spectral resolution, 
to be quite fragmented with density concentrations 
isolated in velocity (and therefore in space) within the 
ejecta.  In this case, rather than representing an 
abundance inhomogeneity or an effect of the optical depth, 
the profiles can best be explained as local knots of 
emission at higher than quenching density.  While the 
density variation is likely small (a factor of 20 to 
50\%), this will need to be properly included in any 
further analysis for the abundance mixing within the ejecta.

\begin{figure}
\plotone{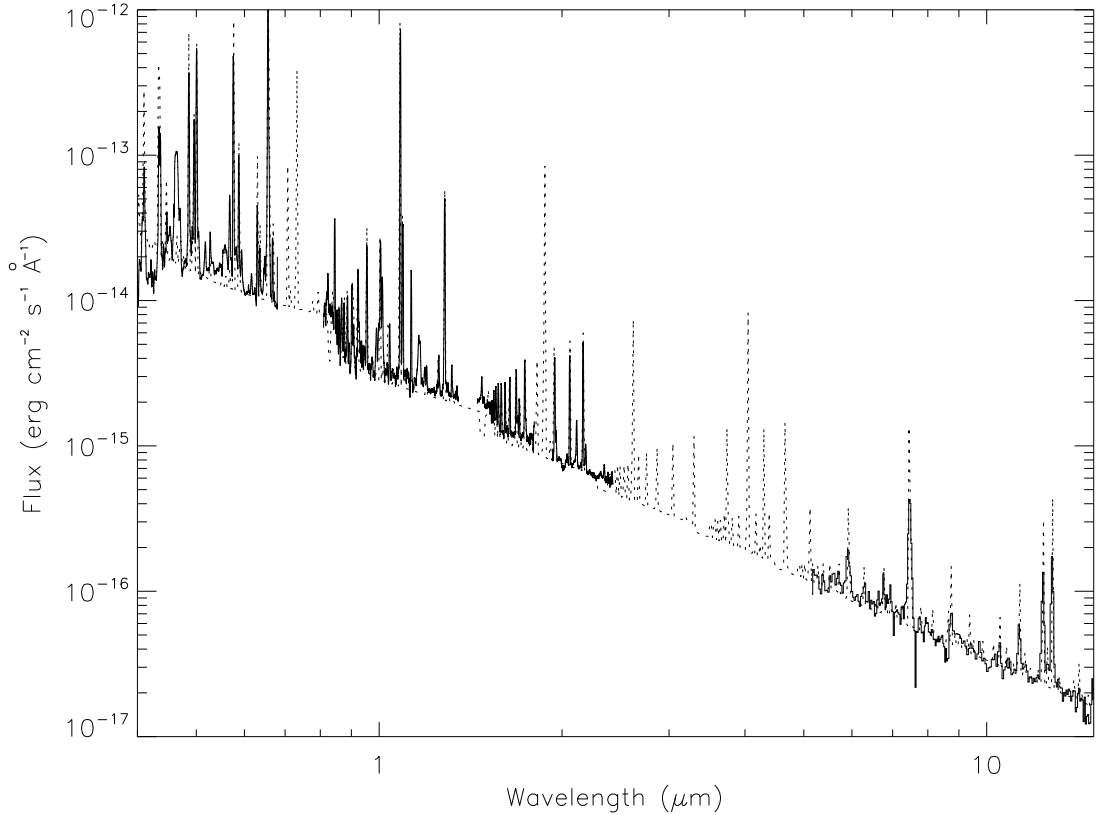}
\caption{Best \cldy\ model (dotted line) fit to the optical, NIR, and 
\textit{Spitzer} spectra (solid lines) obtained during the E2 epoch.  The 
\textit{Spitzer} spectra have been scaled to reflect the $\sim$ 0.7 mag 
difference in the visual light curve between the \textit{Spitzer} and 
optical observations.  The \cldy\ model spectral energy distribution fits
the observations best at a distance of 5.5 kpc.
\label{model-spec}}
\end{figure}

Fitting the \cldy\ model spectral energy distribution to the dereddened
E2 spectra also provides an estimate on the distance.  However, before this
can be done the observed E2 spectra must be placed on a common flux point 
since they were obtained at times when the visual light curve wasn't 
constant.  The visual light curve shows that when the \textit{Spitzer} 
spectra were obtained, V1186 Sco was about 0.7 magnitude brighter 
than when the optical spectrum was taken.  Therefore the relative fluxes of 
the \textit{Spitzer} spectra were reduced by a factor of 1.85.  No relative
flux adjustments were applied to the NIR spectrum since there are no 
light curve measurements at that time.  Fig.~\ref{model-spec} shows the 
model spectral energy distribution which fits the scaled and dereddened 
multiwavelength data assuming a distance of 5.5 kpc.  
This distance determination is farily robust as the model 
fits the emission line ratios and the continuum which are both 
sensitive to the reddening. If the luminosity of the WD was not 
constant during the E2 epoch but rather oscillating during the 
early outburst ({\it e.g.} the super soft source X-ray 
luminosities of \object{V4743 Sgr} \citep{ness03} and \object{RS Oph} 
\citep{os06a,os06b,os06c}) then the derived distance is a lower limit.  As 
a check we apply our \cldy\ luminosity to a different epoch, namely 
visual maximum.  With V$_{max}$ = 9.6, an E(B-V) = 0.45, and assuming a 
bolometric correction of 0 at maximum gives a distance of 
5.4 kpc, consisitent with our previous value. With the uncertainty in 
the reddening, which is likely the largest source of uncertainty, 
we adopt a distance of 5.5$\pm$0.5 kpc and a M$_V$ of $-5.5\pm0.5$ mag 
for V1186 Sco. 

We also attempted to find better fitting \cldy\ models to the E2 spectra with
larger E(B-V) values as indicated by the \citet{schlegel98} reddening map.
Increasing E(B-V) to 0.6 the $\chi^2$ of the best fit model was much higher,
$\sim$75, and the fit of the model continuum to the data was significantly 
worse, particularly at shorter wavelengths.  Assuming the PW Vul M$_V$ of 
$-6.7$, the distance at this reddening is 7.7 kpc.  However, to match the 
dereddened spectra to the best fit model at 7.7 kpc also requires 
invoking a covering factor
\footnote{The fraction ejecta surface area as seen by the photoionizing source.}
of 0.6 which was not required in the low E(B-V)
model.  \cldy\ models fits to spectra dereddened greater than this become 
progressively worse and indicate that our adopted E(B-V) is likely correct.

\section{DISCUSSION}
\label{sec:discus}

The oxygen and nitrogen enhancements predicted by the photionization analysis
are expected from a thermonuclear runaway in the accretion plus WD material
mixture on a CO white dwarf.  Carbon may also be enhanced but the best lines 
to determine this are in the UV.  The slight enhancement of neon is probably 
due to the small amount of $^{22}$Ne in the WD which is produced by helium 
burning while still a giant \citep{L&T94}.  
Nucleosynthesis models that incorporate some $^{22}$Ne in the CO WD 
composition will produce ejecta with slightly enhanced neon abundances 
\citep[see Fig. 1 in][]{J&H98}. The presence of the other 
heavy elements likely reflects that of the secondary 
star since these elements are only created under the extremely high temperature
thermonuclear runaways found on very massive WDs.  

V1186 Sco was the first CO-type nova to reveal the strong [\ion{Ne}{2}] line
common in ONeMg novae using moderate resolution spectroscopic techniques,
but other CO novae may have have also shown this line if observed early enough
in their evolution prior to the condensation of optically thick dust shells. 
\citet{Evans97} reported that [\ion{Ne}{2}] line was weak in the day 396 
spectrum of V705 Cas and may have been present as early as day 251.  The 
[\ion{Ne}{3}] line was also observed in V1425 Aql 650 days after outburst when 
the nova was still in a high ionization state \citep{lyke03}.  \textit{Spitzer} 
observations of future early CO novae spectra will provide additional 
estimates of the amount of neon enhancement.  These results may even 
provide observational clues as to the nature of the mixing mechanism 
and the compositions of CO WDs.

The absolute V band magnitude determined from the photoionization 
analysis is very faint at -5.5$\pm$0.5.  This is consistent with the
results of \citet{Duerbeck81} who found that $<$M$_V>$$ = -6.35\pm0.5$ 
for historic galactic novae of the B-D light curve class.  Our results
imply that V1186 Sco is one of the faintest galactic nova ever discovered.  
In addition, the faint absolute magnitude of V1186 Sco shows that the 
results of MMRD relationships are suspect for novae with irregular light 
curves.

\section{CONCLUSION}
\label{sec:conls}

V1186 Sco was observed with \textit{Spitzer} during two epochs during its 
first year of outburst as part of a Cycle 1 ToO program for classical novae. 
{\bf Its general properties are summarized in Table~\ref{tab:v86props}.}
The \textit{Spitzer} spectra were combined with ground based 
optical and NIR spectra obtained during the same epochs.  The spectra are 
typical of a slowly evolving CO-type nova and unambiguously reveal the
detection of [\ion{Ne}{2}] (12.81 $\mu$m) in this class of outburst, a
highly uncommon event.  The reddening toward V1186 Sco, estimated using 
the intrinsic (B-V) colors and theoretical hydrogen and \ion{O}{1} line 
ratios, converged at E(B-V) = 0.45$\pm$0.1. The well sampled V band 
light curve displayed a slow decline but with many 
secondary peaks that call into question many of the properties derived via 
t$_{2}$ and t$_{3}$ times such as the absolute brightness at maximum.  
The distance derived from the MMRD relationship put V1186 Sco on the 
other side of the Galactic center indicating that the MMRD accuracy is 
suspect for slow novae with irregular light curves.

Photoionization modeling of the data indicates a relatively low mass ejection
event consisting of material enhanced in helium, nitrogen, oxygen and 
some neon.  However, the sulfur, argon and iron abundances must be 
substantially subsolar to explain their lack of prominent lines in the 
observed spectra. The enhanced oxygen and nitrogen, the erratic 
light curve, and the low velocity expansion velocities are all consistent 
with the presence of a CO white dwarf in the system.  The amount of 
neon enhancement (1.3$\pm$0.6 Ne/Ne$_{\odot}$)
is not nearly as high as expected from an outburst on a ONeMg WD.
Rather the neon is likely intrinsic to the WD and mixed with the 
accreted material during the thermonuclear runaway. The distance derived 
from the best fit photoionization model provides a more consistent 
distance of 5.5$\pm$0.5 kpc and a very faint M$_V$ of -5.5$\pm$0.5 mags.

\acknowledgments

We acknowledge with thanks the variable star observations from the AAVSO 
International Database contributed by observers worldwide and used in this 
research. This work is based in part on observations made with the 
{\it{Spitzer}} Space Telescope, which is operated by the Jet Propulsion 
Laboratory, California Institute of Technology, under NASA contract 1407. 
The \textit{Spitzer} Cycle~1 team is supported in part by NASA through 
contracts 1267992 issued by JPL Caltech. This publication also makes use 
of data products from the Two Micron All Sky Survey, which is a joint 
project of the University of Massachusetts and the Infrared Processing and 
Analysis Center/California Institute of Technology, funded by the National 
Aeronautics and Space Administration and the National Science Foundation. 
MFB is grateful to the UK PPARC for the provision of a Senior Fellowship.
SGS acknowledges partial support from NASA, NSF, and \textit{Spitzer}. RDG 
acknowledges support from \textit{Spitzer} contracts 1256406 and 
1215746 issued by JPL Caltech to the University of Minnesota.

{\it Facilities:} \facility{Spitzer (IRS)}, 
\facility{IRTF (SpeX)}, 
\facility{Keck:II (NIRSPEC)}, 
\facility{Bok (B\&C spectrograph)}, 
\facility{MMT (Blue channel spectrograph)},
\facility{Liverpool:2m (RATCam)},
\facility{UKIRT (UIST, CGS4)}

\clearpage

\clearpage


\begin{deluxetable}{lllll}
\tablewidth{0pt}
\tablecaption{LIVERPOOL ROBOTIC 2-m OPTICAL PHOTOMETRY
\label{tab:lrt-phot}}
\tablehead{
\colhead{Date (UT)} & \colhead{MJD} & \colhead{V} &
\colhead{r'} & \colhead{i'}
}
\startdata
12 Aug 2004& 53299.8667& 12.46$\pm$0.03& \nodata       & 10.73$\pm$0.08\\
17 Mar 2005& 53446.2693& 13.37$\pm$0.05& 12.62$\pm$0.04& 13.18$\pm$0.03\\
18 Mar 2005& 53447.2670& 13.25$\pm$0.08& 12.58$\pm$0.04& 13.14$\pm$0.05\\
26 Mar 2005& 53455.2630& 13.43$\pm$0.04& 12.59$\pm$0.04& 13.14$\pm$0.06\\
29 Apr 2005& 53489.1485&    \nodata    & 12.48$\pm$0.12& 13.03$\pm$0.10\\
30 Apr 2005& 53490.1502& 13.47$\pm$0.06& 12.71$\pm$0.05& 13.17$\pm$0.08\\
03 May 2005& 53493.1424& 13.45$\pm$0.05& 12.74$\pm$0.02& 13.23$\pm$0.04\\
18 May 2005& 53508.1003& 13.41$\pm$0.02& 12.94$\pm$0.02& 13.34$\pm$0.25\\
20 May 2005& 53510.0965& 13.46$\pm$0.08& 12.91$\pm$0.05& 13.48$\pm$0.03\\
28 May 2005& 53518.0159& 13.53$\pm$0.05& 12.80$\pm$0.04& 13.37$\pm$0.09\\
29 May 2005& 53519.0148& 13.43$\pm$0.07& 12.83$\pm$0.11& 13.40$\pm$0.09\\
\enddata
\end{deluxetable}

\begin{deluxetable}{llcccc}
\tablewidth{0pt}
\tablecaption{SPECTROSCOPIC OBSERVATION LOG
\label{tab:spec-obslog}}
\tablecolumns{6}
\tablehead{
\colhead{Obs. Date} & \colhead{Age\tablenotemark{a}} & \colhead{MJD} & 
\colhead{Telescope} & \colhead{Instrument} & \colhead{Wavelength} \\
\colhead{(UT)} & \colhead{(d)} & \colhead{(d)} & \colhead{} &
\colhead{} & \colhead{($\mu$m)}
}
\startdata
\cutinhead{E1} 
22 Aug 2004  & 50.1 & 53239.765 & Keck II& NIRSPEC (N-1) &  0.93-1.14\\
23 Aug 2004  & 51.1 & 53240.733 & Keck II& NIRSPEC (K$_l$) & 2.15-2.57 \\
23 Aug 2004  & 51.1 & 53240.758 & Keck II& NIRSPEC (K$_s$) & 1.82-2.24 \\
23 Aug 2004  & 51.1 & 53240.767 & Keck II& NIRSPEC (J) & 1.14-1.42 \\
27 Aug 2004  & 55.5 & 53244.489 & \textit{Spitzer} & IRS & 5-35 \\
29 Aug 2004  & 57.3 & 53246.331 & \textit{Spitzer} & IRS & 5-35 \\
31 Aug 2004  & 59.3 & 53248.310 & \textit{Spitzer} & IRS & 5-35 \\
02 Sept 2004 & 61.9 & 53250.854 & IRTF & SpeX & 0.85-2.5 \\
03 Sept 2004 & 62.1 & 53251.734 & UKIRT & UIST (L) & 2.92-3.64 \\
03 Sept 2004 & 62.1 & 53251.756 & UKIRT & UIST (L') & 3.52-4.15 \\
03 Sept 2004 & 62.1 & 53251.776 & UKIRT & UIST (HK) & 1.40-2.50 \\
04 Sept 2004 & 63.1 & 53252.724 & UKIRT & UIST (M) & 4.46-5.23 \\
04 Sept 2004 & 63.1 & 53252.769 & UKIRT & CGS4 (Z) & 0.80-1.10 \\
04 Sept 2004 & 63.1 & 53252.792 & UKIRT & CGS4 (J) & 1.02-1.34 \\
23 Sept 2004 & 82.1 & 53271.100 & MMT & Blue Channel Spectrograph & 0.34-0.86 \\
\cutinhead{E2} 
21 Mar 2005  & 261.2 & 53450.221 & \textit{Spitzer} & IRS & 5-35 \\
12 Apr 2005  & 283.5 & 53472.470 & Bok 90" & B\&C Spectrograph & 0.35-0.68 \\
16 July 2005 & 378.3 & 53567.266 & IRTF & SpeX & 0.85-2.5 \\
\enddata
\tablenotetext{a}{From discovery date 03 July 2004~UT.}
\end{deluxetable}                                           

\begin{deluxetable}{lcl}
\tablewidth{0pt}
\tablecolumns{3}
\tablecaption{THE NIR/SPITZER LINE FLUXES OF THE MOST PROMINENT FIRST EPOCH (2004)
\label{tab:fepoch1}}
\tablehead{
\colhead{Ion} & \colhead{Wavelength} & \colhead{Flux\tablenotemark{a}} \\
\colhead{} & \colhead{($\mu$m)} & \colhead{(erg s$^{-1}$ cm$^{-2}$)}
}
\startdata
\cutinhead{Near-IR} \\
\ion{O}{1} & 0.8448 & (41.1$\pm$0.2)$\times$10$^{-12}$ \\
\ion{H}{1} (9-3) & 0.9226 & (7.8$\pm$0.1)$\times$10$^{-12}$ \\
\ion{N}{1} &0.9393 & (66.7$\pm$0.2)$\times$10$^{-13}$ \\
\ion{H}{1} (8-3) & 0.9545 & (45.2$\pm$0.3)$\times$10$^{-13}$ \\
Pa$\delta$ & 1.005 & (4.6$\pm$0.1)$\times$10$^{-12}$ \\
$[$\ion{N}{1}$]$ & 1.040 & (57.8$\pm$0.2)$\times$10$^{-13}$ \\
\ion{He}{1} & 1.083 & (56.0$\pm$0.3)$\times$10$^{-13}$ \\
Pa$\gamma$ & 1.094 & (76.4$\pm$0.2)$\times$10$^{-13}$ \\
\ion{O}{1} & 1.129 & (45.8$\pm$0.1)$\times$10$^{-12}$ \\
\ion{He}{1} & 1.254 & (5.5$\pm$0.3)$\times$10$^{-13}$ \\
Pa$\beta$ & 1.282 & (16.0$\pm$0.1)$\times$10$^{-12}$ \\
\ion{O}{1} & 1.317 & (16.8$\pm$0.3)$\times$10$^{-13}$ \\
\ion{H}{1} (11-4) & 1.681 & (10.4$\pm$0.1)$\times$10$^{-13}$ \\
\ion{He}{1} & 1.687 & (12.2$\pm$0.1)$\times$10$^{-13}$ \\
\ion{H}{1} (10-4) & 1.737 & (15.9$\pm$0.2)$\times$10$^{-13}$ \\
\ion{H}{1} (9-4) & 1.818 & (17.5$\pm$0.1)$\times$10$^{-13}$ \\
Pa$\alpha$ & 1.876 & (24.5$\pm$0.1)$\times$10$^{-12}$ \\
Br$\delta$ & 1.945 & (2.8$\pm$0.2)$\times$10$^{-13}$ \\
\ion{He}{1} & 2.059 & (6.6$\pm$0.3)$\times$10$^{-13}$ \\
\ion{He}{1} & 2.113 & (1.1$\pm$0.3)$\times$10$^{-13}$ \\
Br$\gamma$ & 2.166 & (26.8$\pm$0.2)$\times$10$^{-13}$ \\
\ion{He}{1} & 3.086 & (7.9$\pm$0.4)$\times$10$^{-13}$ \\
\ion{He}{1} & 3.331 & (14.2$\pm$0.4)$\times$10$^{-13}$ \\
Pf$\gamma$\tablenotemark{b} & 3.750 & (14.0$\pm$0.4)$\times$10$^{-13}$\\
Br$\alpha$\tablenotemark{b} & 4.052 & (75.6$\pm$0.1)$\times$10$^{-13}$\\
$[$\ion{N}{1}$]$\tablenotemark{b} & 4.650 & (12.9$\pm$0.1)$\times$10$^{-13}$\\
\cutinhead{Spitzer} \\
Hu$\gamma$ & 5.91 & (1.5$\pm$1.2)$\times$10$^{-13}$ \\
\ion{H}{1} (12-7) & 6.77 & (8.3$\pm$6.4)$\times$10$^{-14}$ \\
Pf$\alpha$+Hu$\beta$ & 7.46 & (6.2$\pm$4.6)$\times$10$^{-13}$ \\
\ion{H}{1} (10-7) & 8.76 & (6.1$\pm$2.2)$\times$10$^{-14}$ \\
\ion{H}{1} (12-8) & 10.50 & (1.5$\pm$0.2)$\times$10$^{-13}$ \\
\ion{H}{1} (9-7) & 11.31 & (4.16$\pm$0.2)$\times$10$^{-13}$ \\
Hu$\alpha$ & 12.37 & (1.1$\pm$0.03)$\times$10$^{-12}$ \\
$[$\ion{Ne}{2}$]$ & 12.81 & (3.6$\pm$1.4)$\times$10$^{-14}$ \\
\ion{H}{1} (13-9) & 14.18 & (7.4$\pm$1.4)$\times$10$^{-14}$ \\
\ion{H}{1} (10-8) & 16.21 & (2.1$\pm$0.3)$\times$10$^{-13}$ \\
\ion{H}{1} (8-7) & 19.06 & (3.7$\pm$0.4)$\times$10$^{-13}$ \\
\ion{H}{1} (13-10) & 22.33 & (2.3$\pm$0.2)$\times$10$^{-13}$ \\
\ion{H}{1} (9-8)   & 27.80 & (2.7$\pm$0.2)$\times$10$^{-13}$ \\
\enddata
\tablenotetext{a}{Observed integrated line flux. Not corrected for reddening.}
\tablenotetext{b}{Fluxes are weighted average as line profiles are 
double peaked.}
\end{deluxetable}

\begin{deluxetable}{lrl}
\tablewidth{0pt}
\tablecolumns{3}
\tablecaption{THE LINE FLUXES OF THE MOST PROMINENT LINES IN THE SECOND EPOCH (2005)
\label{tab:fepoch2}}
\tablehead{
\colhead{Ion} & \colhead{Wavelength} & \colhead{Flux\tablenotemark{a}} \\
\colhead{} & \colhead{($\mu$m)} & \colhead{(erg s$^{-1}$ cm$^{-2}$)}
}
\startdata
\cutinhead{Optical} 
H$\delta$ & 0.4100 & (2.6$\pm$0.4)$\times$10$^{-13}$ \\
H$\gamma$ & 0.4340 & (4.4$\pm$0.7)$\times$10$^{-13}$ \\
$[$\ion{O}{3}$]$ & 0.4363 & (3.5$\pm$0.5)$\times$10$^{-13}$ \\
\ion{He}{1} & 0.4471 & (8.0$\pm$2.0)$\times$10$^{-14}$ \\
H$\beta$ & 0.4868 & (1.4$\pm$0.1)$\times$10$^{-12}$ \\
$[$\ion{O}{3}$]$ & 0.4959 & (6.5$\pm$0.7)$\times$10$^{-13}$ \\
$[$\ion{O}{3}$]$ & 0.5007 & (2.4$\pm$0.2)$\times$10$^{-12}$ \\
\ion{N}{2} & 0.5679 & (2.4$\pm$0.7)$\times$10$^{-14}$ \\
$[$\ion{N}{2}$]$ & 0.5755 & (2.9$\pm$0.3)$\times$10$^{-12}$ \\
\ion{He}{1} & 0.5876 & (5.6$\pm$0.8)$\times$10$^{-13}$ \\
$[$\ion{O}{1}$]$ & 0.6300 & (2.5$\pm$0.4)$\times$10$^{-13}$ \\
\ion{He}{1} & 0.6678 & (1.4$\pm$0.2)$\times$10$^{-13}$ \\
\cutinhead{Near-IR} \\
\ion{O}{1} & 0.8448 & (4.1$\pm$0.6)$\times$10$^{-13}$ \\
\ion{H}{1} (10-3) & 0.9015 & (1.5$\pm$0.2)$\times$10$^{-13}$ \\
\ion{H}{1} (9-3) & 0.9229 & (2.4$\pm$0.4)$\times$10$^{-13}$ \\
\ion{H}{1} (8-3)
& 0.9545 & (4.2$\pm$0.6)$\times$10$^{-13}$ \\
Pa$\delta$ & 1.004 & (4.9$\pm$0.7)$\times$10$^{-13}$ \\
$[$\ion{N}{1}$]$ & 1.040 & (8.6$\pm$1.7)$\times$10$^{-14}$ \\
\ion{He}{1} & 1.083 & (1.6$\pm$0.1)$\times$10$^{-11}$ \\
Pa$\gamma$ & 1.094 & (7.6$\pm$0.8)$\times$10$^{-13}$ \\
\ion{O}{1} & 1.129 & (3.0$\pm$0.5)$\times$10$^{-14}$ \\
\ion{He}{1} &1.197 & (2.9$\pm$0.5)$\times$10$^{-14}$ \\
\ion{He}{1} &1.254 & (5.9$\pm$1.0)$\times$10$^{-14}$ \\
Pa$\beta$ & 1.282 & (1.5$\pm$0.2)$\times$10$^{-12}$ \\
\ion{O}{1} & 1.317 & (4.5$\pm$0.9)$\times$10$^{-14}$ \\
\ion{He}{1} & 1.700 & (4.7$\pm$1.0)$\times$10$^{-14}$ \\
Br$\delta$ & 1.945 & (1.8$\pm$0.3)$\times$10$^{-13}$ \\
\ion{He}{1} & 2.058 & (2.1$\pm$0.3)$\times$10$^{-13}$ \\
\ion{He}{1} & 2.113 & (4.9$\pm$1.0)$\times$10$^{-14}$ \\
Br$\gamma$ & 2.166 & (2.9$\pm$0.4)$\times$10$^{-13}$ \\
\cutinhead{Spitzer} \\
Hu$\gamma$ & 5.908 & (1.5$\pm$1.2)$\times$10$^{-13}$ \\
\ion{H}{1} (12-7) & 6.772 & (8.3$\pm$6.4)$\times$10$^{-14}$ \\
Pf$\alpha$+Hu$\beta$ & 7.460 & (6.2$\pm$4.6)$\times$10$^{-13}$ \\
\ion{H}{1} (10-7) & 8.760 & (6.1$\pm$2.2)$\times$10$^{-14}$ \\
\ion{H}{1} (12-8)\tablenotemark{b} & 10.500 & (2.0$\pm$0.6)$\times$10$^{-14}$ \\
\ion{H}{1} (9-7)\tablenotemark{b} & 11.309 & (8.1$\pm$0.7)$\times$10$^{-14}$ \\
Hu$\alpha$\tablenotemark{b} & 12.370 & (2.3$\pm$0.3)$\times$10$^{-13}$ \\
$[$\ion{Ne}{2}$]$\tablenotemark{b} & 12.814 & (3.5$\pm$0.1)$\times$10$^{-13}$ \\
\ion{H}{1} (13-9)\tablenotemark{b} & 14.183 & (2.2$\pm$0.6)$\times$10$^{-14}$ \\
$[$\ion{Ne}{3}$]$ & 15.560 & (9.9$\pm$0.5)$\times$10$^{-14}$ \\
\ion{H}{1} (10-8) & 16.210 & (4.9$\pm$1.4)$\times$10$^{-14}$ \\
\ion{H}{1} (8-7) & 19.060 & (9.3$\pm$1.2)$\times$10$^{-14}$ \\
\ion{H}{1} (13-10)& 22.330 & (5.3$\pm$0.6)$\times$10$^{-14}$ \\
\ion{H}{1} (9-8)  & 27.800 & (5.2$\pm$0.5)$\times$10$^{-14}$ \\
\enddata
\tablenotetext{a}{Observed integrated line flux. Not corrected for reddening.}
\tablenotetext{b}{Weighted average of the SL and SH spectra values.}
\end{deluxetable}

\begin{deluxetable}{lrrrr}
\tablewidth{0pt}
\tablecaption{BEST \cldy\ MODEL FITS TO OBSERVED E2 LINE RATIOS
\label{tab:suitlines}}
\tablehead{
\colhead{Line ratio} & \colhead{$\lambda$} & \colhead{Obs.} & 
\colhead{\cldy} & \colhead{$\chi^2$}
}
\startdata
\cutinhead{Optical\tablenotemark{a}} \\
H$\delta$ & 0.4102 & 0.24 & 0.31 & 1.5 \\
H$\gamma$ & 0.4340 & 0.38 & 0.52 & 2.2 \\
$[$\ion{O}{3}$]$ & 0.4363 & 0.30 & 0.39 & 1.4 \\
\ion{He}{1} & 0.4471 & 0.07 & 0.07 & 0.0 \\
H$\beta$ & 0.4861 & 1.00 & 1.00 & 0.0 \\
$[$\ion{O}{3}$]$ & 0.4959 & 0.44 & 0.26 & 4.0 \\
$[$\ion{O}{3}$]$ & 0.5007 & 1.62 & 0.79 & 6.5 \\
\ion{N}{2} & 0.5679 & 0.12 & 0.01 & 5.5 \\
$[$\ion{N}{2}$]$ & 0.5755 & 1.54 & 1.43 & 0.1 \\
\ion{He}{1} & 0.5876 & 0.29 & 0.20 & 1.6 \\
$[$\ion{O}{1}$]$ & 0.6300 & 0.11 & 0.14 & 0.8  \\
$[$\ion{O}{1}$]$ & 0.6363 & 0.05 & 0.04 & 0.1 \\
H$\alpha$+$[$\ion{N}{2}$]$ & 0.6563 & 4.91 & 4.47 & 0.2 \\
\ion{He}{1} & 0.6678 & 0.06 & 0.05 & 0.5 \\
\cutinhead{Near-IR\tablenotemark{b}} \\
\ion{H}{1} (10-3) & 0.9015 & 0.13 & 0.13 & 0.0 \\
\ion{H}{1} (9-3) & 0.9229 & 0.20 & 0.18 & 0.2 \\
\ion{H}{1} (8-3)
& 0.9546 & 0.35 & 0.23 & 1.8   \\
Pa$\delta$ &1.004 & 0.39 & 0.35 & 0.2 \\
$[$\ion{N}{1}$]$ &1.0400 & 0.07 & 0.05 & 0.5 \\
\ion{He}{1} &1.083 &12.14 & 13.11 & 0.3 \\
Pa$\gamma$ &1.094 & 0.57 & 0.55 & 0.0 \\
\ion{He}{1} &1.197 & 0.02 & 0.02 & 0.8 \\
\ion{He}{1} &1.254 & 0.04 & 0.03 & 0.2 \\
Pa$\beta$ &1.282 & 1.00 & 1.00 & 0.0 \\
\ion{He}{1} &1.700 & 0.03 & 0.02 & 0.3 \\
Br$\delta$ &1.945 & 0.10 & 0.11 & 0.1 \\
\ion{He}{1} &1.954 & 0.01 & 0.02 & 1.7 \\
\ion{He}{1} &2.060 & 0.12 & 0.14 & 0.3 \\
\ion{He}{1} &2.112 & 0.03 & 0.01 & 4.5 \\
Br$\gamma$ &2.166 & 0.16 & 0.16 & 0.0 \\
\cutinhead{Spitzer\tablenotemark{c}} \\
Hu$\gamma$ & 5.907 & 0.65 & 0.50 & 1.3 \\
Pf$\alpha$+Hu$\beta$& 7.458 & 2.70 & 2.75 & 0.0 \\
\ion{H}{1} (9-7)&11.310 & 0.35 & 0.30 & 0.5 \\
Hu$\alpha$&12.370 & 1.00 & 1.00 & 0.0 \\
$[$\ion{Ne}{2}$]$&12.810 & 1.50 & 1.60 & 0.2 \\
$[$\ion{Ne}{3}$]$&15.550 & 0.43 & 0.40 & 0.2 \\
\ion{H}{1} (8-7)&19.060 & 0.40 & 0.40 & 0.0 \\
\ion{H}{1} (9-8)&27.800 & 0.22 & 0.20 & 0.2 \\ 
\enddata
\tablenotetext{a}{Relative to H$\beta$}
\tablenotetext{b}{Relative to Pa$\beta$}
\tablenotetext{c}{Relative to Hu$\alpha$}
\end{deluxetable}

\begin{deluxetable}{ll}
\tablewidth{0pt}
\tablecaption{BEST \cldy\ MODEL PARAMETERS
\label{tab:cldyparam}}
\tablehead{
\colhead{Parameter} & \colhead{Value}
}
\startdata
T$_{BB}$ & 47,000 K \\
Source luminosity & 6.3$\times$10$^{36}$ erg s$^{-1}$ \\
Hydrogen density & 3.2$\times$10$^7$ cm$^{-3}$ \\
Inner radius\tablenotemark{a,b} & 1.1$\times$10$^{15}$ cm \\
Outer radius\tablenotemark{a,b} & 2.2$\times$10$^{15}$ cm \\
Filling factor\tablenotemark{b} & 0.2 \\
He/He$_{\sun}$\tablenotemark{c} & 1.1$\pm$0.3 (10) \\
N/N$_{\sun}$\tablenotemark{c} & 49$^{+34}_{-23}$ (5) \\
O/O$_{\sun}$\tablenotemark{c} & 5.1$^{+0.4}_{-2.8}$ (5) \\
Ne/Ne$_{\sun}$\tablenotemark{c} & 1.3$\pm$0.6 (2) \\
Ejected Mass & 6.5$\times$10$^{-5}$ M$_{\sun}$ \\
Degrees of Freedom & 28 \\
Total $\chi^2$ & 38.0 \\
Reduced $\chi^2$ & 1.36 \\
\enddata
\tablecomments{Model had a hydrogen power law of -3, n(H) $\propto$ r$^{-3}$.}
\tablenotetext{a}{Calculated assuming an inner and outer expansion velocity
of 500 and 1000 km s$^{-1}$, respectively, over 260 days.}
\tablenotetext{b}{Not a free parameter in the model.}
\tablenotetext{c}{Where Log(Solar number abundances relative to hydrogen)
He:-1.0 C: -3.61 N: -4.22 O: -3.34 Ne: -3.93 Mg: -4.47 Al: -5.63
Si: -4.49 S: -4.86 Fe: -4.55 (where all abundances taken from \citet{Asplund05}
except for Ne which is from \citet{GN93}).  All elements not in the 
table were set to solar values.  The number in the parentheses indicates the 
number of \cldy\ lines used in the analysis.}
\end{deluxetable}

\begin{deluxetable}{ll}
\tablewidth{0pt}
\tablecaption{V1186 Sco PROPERTIES \label{tab:v86props}}
\tablehead{
\colhead{} & \colhead{}
}
\startdata
Equatorial Coordinates & RA=17:12:51.3 Dec=-30:56:39 (J2000) \\
Galactic Coordinates & $l$=354.4 $b$=4.8 (G2000) \\
Date Discovered & 3.146 July 2004 (MJD=53189.146) \\
Visual Maximum & 9 July 2004 \\
V$_{max}$ & 9.6 \\
t$_2$ and t$_3$ & 20 and 70 days \\
Line widths & $\sim$ 500-1000 km s$^{-1}$ \\
E(B-V) & 0.45$\pm$0.10 \\
M$_{V}^{max}$ & -5.5$\pm$0.5 \\
Distance & 5.5$\pm$0.5 kpc \\
Dust formation? & None detected \\
\enddata
\end{deluxetable}

\clearpage


\end{document}